\documentclass[final,1p,times]{elsarticle}

\usepackage{graphics}

\usepackage{amssymb}
\usepackage{amsfonts}
\usepackage{subeqn}
\usepackage{mathrsfs}
\usepackage{multirow}
\biboptions{sort&compress}

\journal{}

\begin{document}

\begin{frontmatter}

\title{Maxwell-Stefan theory based lattice Boltzmann model for diffusion in multicomponent mixtures}


\author{Zhenhua Chai\fnref{a,b}}\ead{hustczh@hust.edu.cn}
\author{Xiuya Guo\fnref{a}}
\author{Lei Wang\fnref{c}}
\author{Baochang Shi\corref{cor1}\fnref{a,b}}\cortext[cor1]{Corresponding author at: School of Mathematics and Statistics, Huazhong University of Science and Technology, Wuhan, 430074, China. Tel./fax: +86 27 8754 3231. \hspace*{20pt}}\ead{shibc@hust.edu.cn}
\address[a]{School of Mathematics and Statistics, Huazhong University of Science and Technology, Wuhan, 430074, China}
\address[b]{Hubei Key Laboratory of Engineering Modeling and Scientific Computing, Huazhong University of Science and Technology, Wuhan 430074, China}
\address[c]{School of Mathematics and Physics, China University of Geosciences, Wuhan, 430074, China}
\begin{abstract}

 The phenomena of diffusion in multicomponent (more than two components) mixtures are very universal in both science and engineering, and from mathematical point of view, they are usually described by the Maxwell-Stefan (MS) based continuum equations. In this paper, we propose a multiple-relaxation-time lattice Boltzmann (LB) model for the mass diffusion in multicomponent mixtures, and also perform a Chapman-Enskog analysis to show that the MS based continuum equations can be correctly recovered from the developed LB model. In addition, considering the fact that the MS based continuum equations are just a diffusion type of partial differential equations, we can also adopt much simpler lattice structures to reduce the computational cost of present LB model. We then conduct some simulations to test this model, and find that the results are in good agreement with some available works. Besides, the reverse diffusion, osmotic diffusion and diffusion barrier phenomena are also captured. Finally, compared to the kinetic theory based LB models for multicomponent gas diffusion, the present model does not include any complicated interpolations, and its collision process can be still implemented locally. Therefore, the advantages of single-component LB method can also be preserved in present LB model.

\end{abstract}


\end{frontmatter}

%
%

\section{Introduction}

Diffusion, as one of important transport processes, has received increasing attention for its physical significance in the study of most chemical engineering and energy problems \cite{Taylor1993, Krishna1997, Wesselingh2000, Bao2018}. From the physical point of view, the diffusion is the result of random molecular motion \cite{Taylor1993}, while mathematically, the diffusion can also be depicted by two classic continuum mechanical models \cite{Taylor1993, Krishna1997, Wesselingh2000, Bird2002}, i.e., the Fick's law based equations \cite{Fick1855} and Maxwell-Stefan (MS) theory based equations \cite{Maxwell1867, Stefan1871}. In the first model, the diffusion flux of one component is assumed to be proportional to the negative of its concentration gradient, and the cross effects (or the influences of other components) in a system with more than two components are not included although they are well-known to appear in reality. In the past decades, the Fick's law based continuum equations have been widely used to investigate the multicomponent diffusion problems for their simplicity, while they are only valid for the diffusion in binary mixtures or diffusion of a dilute species in a multicomponent system, and thus some curious phenomena caused by the cross effects in the multicomponent mixtures, including the reverse diffusion (up-hill diffusion in direction of the concentration gradient), osmotic diffusion (diffusion without a concentration gradient) and diffusion barrier (no diffusion with a concentration gradient) \cite{Toor1957, Duncan1962}, cannot be captured by this kind of continuum equations \cite{Taylor1993,Krishna1997, Wesselingh2000}. To account for such complex diffusion phenomena observed in the multicomponent mixtures \cite{Duncan1962, Arnold1967, Krishna1997}, the MS theory based continuum equations must be considered. However, the MS theory based continuum equations are nonlinear coupled partial differential equations, and usually it is difficult or even impossible to obtain their analytical solutions. For this reason, most of available works focus on the approximate solutions of such complicated partial differential equations. Basically, there are two possible ways that can be used to obtain the approximate solutions of the MS theory based continuum equations. The first one is theoretical approach \cite{Taylor1993}. In this approach, usually the MS theory based continuum equations are first linearized where the effective diffusivities are assumed to be constants, then some mathematical methods are applied to derive the solutions of the linearized equations \cite{Toor1964,Stewart1964,Taylor1993}. Although this approach can be used to reveal the complex diffusion mechanisms in multicomponent mixtures \cite{Toor1957,Duncan1962,Arnold1967,Veltzke2015}, it is usually limited to one-dimensional problems, and sometimes may also bring some undesirable errors due to the assumption of composite-independent effective diffusivities in the linearized equations \cite{Weber2016}. The other one is numerical approach \cite{Giovangigli1999,Geiser2015}. In the derivation of approximate solutions with this approach, there are no any assumptions, but we need to develop some numerical schemes with certain truncation errors to discretize MS theory based continuum equations. With the development of computer technology and scientific computing, this approach has become more popular in solving the MS theory based continuum equations. Actually, some numerical methods, including the finite-difference method \cite{III2001,Boudin2012}, finite-volume method \cite{Kumar2010,Mazumder2006,Peerenboom2011,Gandhi2012,Kohoff2012,Woo2018}, finite-element method \cite{Boudin2010,Bottcher2010,McLeod2014} and smoothed particle hydrodynamics method \cite{Hirschler2016}, have been developed to solve MS theory based continuum equations.

As an alternative to above mentioned numerical methods for multicomponent diffusion problems, the lattice Boltzmann (LB) method \cite{Chen1998, Succi2001}, a mesoscopic numerical method developed from lattice gas automata \cite{Frisch1986} or continuum Boltzmann equation \cite{He1997}, has also been adopted to study the diffusion in multicomponent mixtures for its kinetic background \cite{Guo2013, Kruger2017}. Generally, there are two main kinds of LB models, i.e., kinetic theory based LB models \cite{Luo2003,McCracken2005,Xu2005,Guo2005,Asinari2006,Asinari2008,Zheng2010,Arcidiacono2006} and pseudo-potential based LB models \cite{Shan1995,Shan1996,Chai2012}, that have been used for multicomponent diffusion problems. The LB models of first kind are developed from some particular kinetic equations for multicomponent gas mixtures \cite{Guo2013}, and can also be viewed as a natural extension of the LB models for single-component fluid flows. Due to the solid physical background, these LB models have also been applied to investigate the multicomponent gas transport in complex porous media at pore scale \cite{Joshi2007,Asinari2007,Kim2009,Rama2010,Tong2014}. However, the LB models of this kind also have some limitations. For instance, when these LB models are adopted to study the diffusion in multicomponent mixtures with different molecular weights, some interpolations \cite{McCracken2005}, modifications on the equilibrium distribution function of the truncated Maxwellian form \cite{McCracken2005, Asinari2006,Asinari2008,Arcidiacono2006}, or finite-difference techniques \cite{Xu2005,Guo2005,Zheng2010} must be needed. In the LB models of second kind, a so-called pseudo-potential is introduced to depict the interaction between different species \cite{Shan1995,Shan1996}. To obtain correct macroscopic governing equations for multicomponent problems with different molecular weights, however, some modifications on equilibrium distribution functions and more discrete velocities are usually needed, as reported in Ref. \cite{Chai2012}.

Different from the models mentioned above, Hosseini et al. \cite{Hosseini2018} also developed a LB model for multicomponent diffusion, which can be considered as a direct numerical solver to macroscopic governing equations for multicomponent fluid problems. To obtain correct governing equations from their LB model, a more complicated equilibrium distribution function including a gradient term is adopted, and to further determine the gradient term in the equilibrium distribution function, the local schemes developed in the framework of LB method \cite{Chai2013,Chai2014,Meng2016} are used. We would like to point out that, however, to obtain the gradient term related to species $i$ in a multicomponent system, a linear system of equations rather than Eq.~(28) in Ref. \cite{Hosseini2018} must be solved. In this work, we would propose a MS theory based multiple-relaxation-time LB model for diffusion in multicomponent mixtures where a much simpler equilibrium distribution function is adopted. In our model, the coupling effects among different species are reflected through corresponding cross collision terms, which is similar to that in the kinetic models for multicomponent gas mixtures \cite{Chapman1970}. In addition, through the Chapman-Enskog analysis, one can also show that the MS theory based continuum equations can be recovered correctly from this model.

The rest of the paper is organized as follows. In Section 2, the MS theory based continuum equations for diffusion in multicomponent mixtures are first introduced, then the LB model for these equations is developed in Section 3. In Section 4, we present some numerical results and discussion, and finally, some conclusions are given in Section 5.

\section{Maxwell-Stefan theory of the diffusion in multicomponent mixtures}\label{section2}

For an ideal gas mixture composed of $n$ chemical species, the molar concentration $c_{i}$ of species $i$ satisfies the following continuum equation \cite{Taylor1993, Bothe2010, Boudin2012},
\begin{equation}\label{eq2-1}
\partial_{t}c_{i}+\nabla\cdot \mathbf{N}_{i}=0,
\end{equation}
or
\begin{equation}\label{eq2-2}
\partial_{t}c_{i}+\nabla\cdot (c_{i}\mathbf{u})=-\nabla\cdot\mathbf{J}_{i},\ \ 1 \leq i \leq n,
\end{equation}
where $\mathbf{N}_{i}=c_{i}\mathbf{u}_{i}$ is the molar flux of species $i$, $\mathbf{J}_{i}=c_{i}(\mathbf{u}_{i}-\mathbf{u})$ is the molar diffusion flux.
$\mathbf{u}_{i}$ is the molar velocity, $\mathbf{u}$ is the molar average velocity, and is defined by
\begin{equation}\label{eq2-3}
\mathbf{u}=\frac{1}{c_{n}}\sum_{i=1}^{n}{c_{i}\mathbf{u}_{i}}=\sum_{i=1}^{n}{\xi_{i}\mathbf{u}_{i}}, \ \ c_{t}=\sum_{i=1}^{n}{c_{i}},
\end{equation}
where $\xi_{i}=c_{i}/c_{t}$ is the mole fraction of species $i$, $c_{t}$ is total molar concentration, and is a constant at the isobaric condition. Based on the definitions of molar diffusion flux and mole fraction, one can obtain the following relations,
\begin{subequations}
\begin{equation}\label{eq2-3a}
\sum_{i=1}^{n}{\mathbf{J}_{i}}=0,
\end{equation}
\begin{equation}\label{eq2-3b}
\sum_{i=1}^{n}{\xi_{i}}=1.
\end{equation}
\end{subequations}
In the present work, we only consider the diffusion in multicomponent mixtures, and the molar average velocity $\mathbf{u}$ is assumed to be zero. In this case, Eqs.~(\ref{eq2-1}) and (\ref{eq2-2}) can be rewritten as
\begin{equation}\label{eq2-4}
\partial_{t}c_{i}=-\nabla\cdot\mathbf{J}_{i},\ \ 1 \leq i \leq n.
\end{equation}
In the following, we would present some details on how to determine the molar diffusion flux $\mathbf{J}_{i}$ in the framework of MS theory.

In the MS theory, the thermodynamical driving force $\mathbf{d}_{i}$ exerted on species $i$ is balanced by the total friction force of species $i$ and other species. For an ideal gas mixture at the constant pressure $P$ and temperature $T$, the driving force $\mathbf{d}_{i}$ takes the following form \cite{Taylor1993, Krishna1997, Bothe2010},
\begin{equation}\label{eq2-5}
\mathbf{d}_{i}=\frac{\xi_{i}}{RT}\nabla\mu_{i}=\nabla\xi_{i},
\end{equation}
where $\mu_{i}$ is the chemical potential, $R$ is the gas constant.
On the other hand, if the mutual friction force between species $i$ and $j$ is assumed to be proportional to the relative velocity and mole fraction \cite{Krishna1997, Bothe2010}, and based on the balance between the driving forces, one can obtain
\begin{equation}\label{eq2-6}
\nabla\xi_{i}=-\sum_{j\neq i}^{n}\sigma_{ij}\xi_{i}\xi_{j}(\mathbf{u}_{i}-\mathbf{u}_{j})=-\sum_{j\neq i}^{n}\frac{\sigma_{ij}(\xi_{j}c_{i}\mathbf{u}_{i}-\xi_{i}c_{j}\mathbf{u}_{j})}{c_{t}},
\end{equation}
$\sigma_{ij}>0$ is the drag coefficient. Through incorporating the molar diffusion flux $\mathbf{J}_{i}$, we can also rewrite Eq.~(\ref{eq2-6}) as
\begin{equation}\label{eq2-7}
\nabla\xi_{i}=-\sum_{j\neq i}^{n}\frac{\xi_{j}\mathbf{J}_{i}-\xi_{i}\mathbf{J}_{j}}{c_{t}D_{ij}},
\end{equation}
which is the so-called MS equation for species $i$. $D_{ij}=1/\sigma_{ij}$ is the MS diffusivity, and is also symmetric based on the fact $\sigma_{ij}=\sigma_{ji}$ \cite{Taylor1993,Bothe2010}.

Next, we would determine the explicit expression of molar diffusion flux $\mathbf{J}_{i}$ from the MS equations. Through a summation of Eq. (\ref{eq2-7}) over $i$ and with the help of Eq. (\ref{eq2-3b}) or the symmetry of $D_{ij}$, we can first derive the following equation,
\begin{equation}\label{eq2-8}
\sum_{i=1}^{n}\nabla\xi_{i}=-\sum_{i=1}^{n}\sum_{j\neq i}^{n}\frac{\xi_{j}\mathbf{J}_{i}-\xi_{i}\mathbf{J}_{j}}{c_{t}D_{ij}}=0,
\end{equation}
which indicates that the MS equations for all $n$ species are linearly dependent. To eliminate the linear dependence, we can use Eqs.~(\ref{eq2-3a}) to remove the molar diffusion flux $\mathbf{J}_{n}$ and the equation related to $\nabla\xi_{n}$ from the MS equations~(\ref{eq2-7}),
\begin{eqnarray}\label{eq2-8a}
\nabla\xi_{i} & = & -\frac{1}{c_{t}}\big(\mathbf{J}_{i}\sum_{j\neq i}^{n}\frac{\xi_{j}}{D_{ij}}-\xi_{i}\sum_{j\neq i}^{n}\frac{\mathbf{J}_{j}}{D_{ij}}\big)\\ \nonumber
& = &  -\frac{1}{c_{t}}\big(\mathbf{J}_{i}\sum_{j\neq i}^{n}\frac{\xi_{j}}{D_{ij}}-\xi_{i}\sum_{j\neq i}^{n-1}\frac{\mathbf{J}_{j}}{D_{ij}}+\frac{\xi_{i}}{D_{in}}\sum_{j=1}^{n-1}\mathbf{J}_{j}\big)\\ \nonumber
& = & -\frac{1}{c_{t}}\big[\big(\frac{\xi_{i}}{D_{in}}+\sum_{j\neq i}^{n}\frac{\xi_{j}}{D_{ij}}\big)\mathbf{J}_{i}+\xi_{i}\sum_{j\neq i}^{n-1}\big(\frac{1}{D_{in}}-\frac{1}{D_{ij}}\big)\mathbf{J}_{j}\big],\ \ 1\leq i\leq n-1,
\end{eqnarray}
which can also be rewritten in a matrix form,
\begin{equation}\label{eq2-9}
c_{t}\left(\begin{array}{c}
\nabla \xi_{1}  \\
\nabla \xi_{2}  \\
\vdots  \\
\nabla \xi_{n-1}
\end{array} \right)=-\mathbf{B}\left(\begin{array}{c}
\mathbf{J}_{1}  \\
\mathbf{J}_{2}  \\
\vdots  \\
\mathbf{J}_{n-1}
\end{array} \right),
\end{equation}
where $\mathbf{B}$ is a $(n-1)\times(n-1)$ matrix, and the element $B_{ij}$ is given by
\begin{equation}\label{eq2-10}
B_{ij}=\left\{\begin{array}{c} \xi_{i}\big(\frac{1}{D_{in}}-\frac{1}{D_{ij}}\big),\ \ \ \ i\neq j,\\
\frac{\xi_{i}}{D_{in}}+\sum_{k\neq i}^{n}\frac{\xi_{k}}{D_{ik}}, \ \ i=j.
\end{array}\right.
\end{equation}
If the matrix $\mathbf{B}$ is assumed to be invertible, Eq.~(\ref{eq2-9}) can be written as
\begin{equation}\label{eq2-11}
\left(\begin{array}{c}
\mathbf{J}_{1}  \\
\mathbf{J}_{2}  \\
\vdots  \\
\mathbf{J}_{n-1}
\end{array} \right)=-c_{t}\tilde{\mathbf{D}}\left(\begin{array}{c}
\nabla \xi_{1}  \\
\nabla \xi_{2}  \\
\vdots  \\
\nabla \xi_{n-1}
\end{array} \right),
\end{equation}
where $\tilde{\mathbf{D}}=\mathbf{B}^{-1}$ is the matrix of the effective diffusivity or generalized Fick diffusivity, and is also a function of $D_{ij}$ and $\xi_{i}$. From Eq.~(\ref{eq2-11}), we can express the molar diffusion flux $\mathbf{J}_{i}$ as
\begin{equation}\label{eq2-12}
\mathbf{J}_{i}=-c_{t}\sum_{j=1}^{n-1}\tilde{D}_{ij}\nabla\xi_{j}, \ \ 1\leq i\leq n-1,
\end{equation}
which can be considered as the generalized Fick's law.

Substituting Eq.~(\ref{eq2-12}) into Eq.~(\ref{eq2-4}) and based on the definition of mole fraction, one can obtain the MS theory based continuum equations,
\begin{equation}\label{eq2-13}
\partial_{t}\xi_{i}=\nabla\cdot\big(\sum_{j=1}^{n-1}\tilde{D}_{ij}\nabla\xi_{j}\big), \ \ 1\leq i\leq n-1,
\end{equation}
while for the species $n$, the mole fraction $\xi_{n}$ can be determined directly by Eq.~(\ref{eq2-3b}). In the following, we would adopt the MS based continuum equations~(\ref{eq2-13}) to study the diffusion in multicomponent mixtures.

Here we would also like to present some remarks on the matrix $\tilde{\mathbf{D}}$ and molar diffusion flux $\mathbf{J}_{i}$ [Eq.~(\ref{eq2-12})].

\noindent
\textbf{Remark I}: For a binary mixture, we can obtain the following relations,
\begin{equation}\label{eq2-14}
\mathbf{J}_{1}=-\mathbf{J}_{2},\ \ \xi_{1}+\xi_{2}=1,\ \  D_{12}=D_{21}=\mathcal{D},
\end{equation}
where $\mathcal{D}$ is denoted as the diffusivity. Based on Eq.~(\ref{eq2-14}), one can rewrite the molar diffusion flux $\mathbf{J}_{i}$, i.e., Eq.~(\ref{eq2-12}), as
\begin{equation}\label{eq2-12a}
\mathbf{J}_{i}=-c_{t}\mathcal{D}\nabla\xi_{i}, \ \ i=1,\ 2,
\end{equation}
which is just the classic Fick's law.

\noindent
\textbf{Remark II}: For a multicomponent system where all MS diffusivities $D_{ij}\ (1\leq i,j \leq n-1)$ are equal to each other, and are represented by $\mathcal{D}$, the matrix $\mathbf{B}$ [see Eq.~(\ref{eq2-10})] and its inverse $\tilde{\mathbf{D}}$ can be simplified by
\begin{equation}\label{eq2-15}
\mathbf{B}=\frac{1}{\mathcal{D}}\mathbf{I}, \ \ \tilde{\mathbf{D}}=\mathcal{D}\mathbf{I},
\end{equation}
where $\mathbf{I}$ is the $(n-1)\times(n-1)$ unit matrix. With the help of Eq.~(\ref{eq2-15}), the molar diffusion flux $\mathbf{J}_{i}$ [see Eq.~(\ref{eq2-12})] can be written in a much simpler form,
\begin{equation}\label{eq2-12b}
\mathbf{J}_{i}=-c_{t}\mathcal{D}\nabla\xi_{i}, \ \ 1\leq i\leq n-1,
\end{equation}
which is also consistent with the Fick's law.

\noindent
\textbf{Remark III}: For a multicomponent system where the species $i$ is dilute ($\xi_{i}\rightarrow0$), the matrix $\mathbf{B}$ and its inverse $\tilde{\mathbf{D}}$ can be approximated by
\begin{equation}\label{eq2-16}
\mathbf{B}=\frac{1}{\mathcal{D}_{eff}}\mathbf{I}, \ \ \tilde{\mathbf{D}}=\mathcal{D}_{eff}\mathbf{I},
\end{equation}
where $\mathcal{D}_{eff}$ is the effective diffusivity, and is defined by
\begin{equation}\label{eq2-17}
\mathcal{D}_{eff}=1/\sum_{k\neq i}^{n}\frac{\xi_{k}}{D_{ik}}.
\end{equation}
Then we can obtain the molar diffusion flux $\mathbf{J}_{i}$.
\begin{equation}\label{eq2-12c}
\mathbf{J}_{i}=-c_{t}\mathcal{D}_{eff}\nabla\xi_{i},
\end{equation}
It is clear that Eq.~(\ref{eq2-12c}) is similar to the Fick's law, but the effective diffusivity $\mathcal{D}_{eff}$ is a function of $\xi_{j}\ (j\neq i)$ and $D_{ij}$ rather than a constant.

\noindent
\textbf{Remark IV}: For a three-component system, the matrix $\mathbf{B}$ can be explicitly expressed as
\begin{equation}
\mathbf{B}=\left(\begin{array}{cc}
\frac{1}{D_{13}}+\xi_{2}\big(\frac{1}{D_{12}}-\frac{1}{D_{13}}\big) & \xi_{1}\big(\frac{1}{D_{13}}-\frac{1}{D_{12}}\big) \\
\xi_{2}\big(\frac{1}{D_{23}}-\frac{1}{D_{12}}\big) & \frac{1}{D_{23}}+\xi_{1}\big(\frac{1}{D_{12}}-\frac{1}{D_{23}}\big)
\end{array} \right),
\end{equation}
where Eq.~(\ref{eq2-3b}) has been used. It is also easy to show that the matrix $\mathbf{B}$ is invertible since its determinant is not equal to zero, as seen below,
\begin{equation}\label{eq2-18}
|\mathbf{B}|=\frac{\xi_{1}}{D_{12}D_{13}}+\frac{\xi_{2}}{D_{12}D_{23}}+\frac{\xi_{3}}{D_{13}D_{23}}>0.
\end{equation}
Then one can obtain the inverse of matrix $\mathbf{B}$,
\begin{equation}\label{eq2-19}
\tilde{\mathbf{D}}=\frac{1}{\frac{\xi_{1}}{D_{12}D_{13}}+\frac{\xi_{2}}{D_{12}D_{23}}+\frac{\xi_{3}}{D_{13}D_{23}}}\left(\begin{array}{cc}
\frac{1}{D_{23}}+\xi_{1}\big(\frac{1}{D_{12}}-\frac{1}{D_{23}}\big) & \xi_{1}\big(\frac{1}{D_{12}}-\frac{1}{D_{13}}\big) \\
\xi_{2}\big(\frac{1}{D_{12}}-\frac{1}{D_{23}}\big) & \frac{1}{D_{13}}+\xi_{2}\big(\frac{1}{D_{12}}-\frac{1}{D_{13}}\big)
\end{array} \right),
\end{equation}
or equivalently,
\begin{equation}\label{eq2-19}
\tilde{\mathbf{D}}=\frac{1}{\xi_{1}D_{23}+\xi_{2}D_{13}+\xi_{3}D_{12}}\left(\begin{array}{cc}
D_{13}\big(\xi_{1}D_{23}+(1-\xi_{1})D_{12}\big) & \xi_{1}D_{23}\big(D_{13}-D_{12}\big) \\
\xi_{2}D_{13}\big(D_{23}-D_{12}\big) & D_{23}\big(\xi_{2}D_{13}+(1-\xi_{2})D_{12}\big)
\end{array} \right),
\end{equation}
which can be used to determine the explicit expression of molar diffusion flux $\mathbf{J}_{i}$.

\section{Lattice Boltzmann model for Maxwell-Stefan theory based continuum equations}\label{section3}

The LB method, as one of kinetic theory based numerical approaches, has made a great progress in the study of complex fluid flows in the past three decades \cite{Chen1998,Succi2001,Guo2013,Kruger2017,Chen2014,Huang2015,Li2016,Liu2016,Xu2017}, while simultaneously, it can also be considered as a general solver to nonlinear diffusion and convection-diffusion equations \cite{Dawson1993,Wolf-Gladrow1995,Ginzburg2005a,Chopard2009,Shi2009,Huber2010,Yoshida2010,Chai2013,Li2013,Chai2016,Aursjo2017,Chai2018}. Based on the collision term, the basic LB models can be classified into three categories, i.e., the single-relaxation-time (SRT) LB model (or lattice BGK model) \cite{Qian1992}, the two-relaxation-time (TRT) LB model \cite{Ginzburg2008}, and the multiple-relaxation-time (MRT) LB model (or generalized LB model) \cite{dHumieres1992}. Here we would consider the MRT-LB model for its advantages in generalization, stability and accuracy \cite{Lallemand2000,Luo2011,Cui2016,Pan2006,Chai2016b}.

Considering the fact that the MS theory based continuum equations (\ref{eq2-13}) are only a special case of nonlinear coupled diffusion equations, some available LB models for diffusion or convection-diffusion equations can be extended to solve the MS theory based continuum equations. In this work, we would propose a MRT-LB model for these continuum equations, and incorporate the cross collision terms in this model to reflect the coupling effects among different species which is similar to that in Ref. \cite{Huber2010}.

\subsection{Multiple-relaxation-time lattice Boltzmann model}

In the MRT-LB model for Eq.~(\ref{eq2-13}), the evolution equation can be written as \cite{dHumieres1992,Lallemand2000}
\begin{eqnarray}\label{eq3-2}
f_{k}^{i}(\mathbf{x} + \mathbf{c}_{k}\delta t,\;t + \delta t) = f_{k}^{i}(\mathbf{x},\;t)-\sum_{j=1}^{n-1}(\mathbf{M}^{-1}\mathbf{\Lambda}^{ij}\mathbf{M})_{k\alpha}\big[f_{\alpha}^{j}(\mathbf{x},\;t) - f_{\alpha}^{j,(eq)}(\mathbf{x},\;t)\big],\ \ 1\leq i\leq n-1,
\end{eqnarray}
where $f_{k}^{i}(\mathbf{x},\;t)$ ($k=0,\ 1,\ \cdots,\ q-1$ or $1,\; 2\; \cdots,\ q$ with $q$ representing the number of discrete velocity directions) is the distribution function of species $i$ at position $\mathbf{x}$ and time $t$, $\mathbf{c}_{k}$ is the discrete velocity. $f_{k}^{i,(eq)}(\mathbf{x},\;t)$ is the equilibrium distribution function, and for diffusion problems, it can be simply given by \cite{Wolf-Gladrow1995,Chai2016,Chai2018}
\begin{equation}\label{eq3-3}
f_{k}^{i,(eq)}(\mathbf{x},\;t)=\omega_{k}\xi_{i},
\end{equation}
where $\omega_{k}$ is the weight coefficient. In some commonly used D$d$Q$q$ ($q$ velocity directions in $d$ dimensional space) lattice models, the weight coefficient $\omega_{k}$ and discrete velocity $\mathbf{c}_{k}$ are defined as \cite{Chai2018}

\begin{subequations}\label{eq3-4}
\textrm{D$1$Q$2$: $\;$}
\begin{equation}
\omega_{1}=\omega_{2}=\frac{1}{2},
\end{equation}
\begin{equation}
\mathbf{c}=(1, -1)c,
\end{equation}
\end{subequations}

\begin{subequations}\label{eq3-5}
\textrm{D$1$Q$3$: $\;$}
\begin{equation}
\omega_{0}=\frac{2}{3}, \; \omega_{1}=\omega_{2}=\frac{1}{6}
\end{equation}
\begin{equation}
\mathbf{c}=(0, 1, -1)c,
\end{equation}
\end{subequations}

\begin{subequations}\label{eq3-6}
\textrm{D$2$Q$4$: $\;$}
\begin{equation}
\omega_{i=1-4}=\frac{1}{4},
\end{equation}
\begin{equation}
\mathbf{c}=\left(\begin{array}{cccc}
1 & 0 & -1 & 0 \\
0 & 1 & 0 & -1
\end{array} \right)c,
\end{equation}
\end{subequations}

\begin{subequations}\label{eq3-7}
\textrm{D$2$Q$5$: $\;$}
\begin{equation}
\omega_{0}=\frac{1}{3}, \; \omega_{i=1-4}=\frac{1}{6},
\end{equation}
\begin{equation}
\mathbf{c}=\left(\begin{array}{ccccc}
0 & 1 & 0 & -1 & 0 \\
0 & 0 & 1 & 0 & -1
\end{array} \right)c,
\end{equation}
\end{subequations}

\begin{subequations}\label{eq3-8}
\textrm{D$2$Q$9$: $\;$}
\begin{equation}
\omega_{0}=\frac{4}{9}, \; \omega_{i=1-4}=\frac{1}{9}, \; \omega_{i=5-8}=\frac{1}{36},
\end{equation}
\begin{equation}
\mathbf{c}=\left(\begin{array}{ccccccccc}
0 & 1 & 0 & -1 & 0 & 1 & -1 & -1 & 1 \\
0 & 0 & 1 & 0 & -1 & 1 & 1 &-1 & -1
\end{array} \right)c,
\end{equation}
\end{subequations}

\begin{subequations}\label{eq3-9}
\textrm{D$3$Q$6$: $\;$}
\begin{equation}
\omega_{i=1-6}=\frac{1}{6},
\end{equation}
\begin{equation}
\mathbf{c}=\left(\begin{array}{cccccc}
1 & -1 & 0 & 0 & 0 & 0 \\
0 & 0 & 1 & -1 & 0 & 0\\
0 & 0 & 0 & 0 & 1 & -1
\end{array} \right)c,
\end{equation}
\end{subequations}

\begin{subequations}\label{eq3-10}
\textrm{D$3$Q$7$: $\;$}
\begin{equation}
\omega_{0}=\frac{1}{4}, \; \omega_{i=1-6}=\frac{1}{8},
\end{equation}
\begin{equation}
\mathbf{c}=\left(\begin{array}{ccccccc}
0 & 1 & -1 & 0 & 0 & 0 & 0 \\
0 & 0 & 0 & 1 & -1 & 0 & 0\\
0 & 0 & 0 & 0 & 0 & 1 & -1
\end{array} \right)c,
\end{equation}
\end{subequations}

\begin{subequations}\label{eq3-11}
\textrm{D$3$Q$15$: $\;$}
\begin{equation}
\omega_{0}=\frac{2}{9}, \; \omega_{i=1-6}=\frac{1}{9}, \; \omega_{i=7-14}=\frac{1}{72},
\end{equation}
\begin{equation}
\mathbf{c}=\left(\begin{array}{ccccccccccccccc}
0 & 1 & -1 & 0 & 0 & 0 & 0 & 1 & 1 & 1 & 1 & -1 & -1 & -1 & -1\\
0 & 0 & 0 & 1 & -1 & 0 & 0 & 1 & -1 & -1 & 1 & 1 & -1 & -1 & 1\\
0 & 0 & 0 & 0 & 0 & 1 & -1 & 1 & -1 & 1 & -1 & 1 & -1 & 1 & -1
\end{array} \right)c.
\end{equation}
\end{subequations}
$c=\delta x/\delta t$ is the lattice speed, $\delta x$ and $\delta t$ are the lattice spacing and time step, respectively. We note that although there are some other lattice models \cite{Guo2013,Kruger2017}, for the sake of brevity, they are not presented here. $\mathbf{M}$ is a $q\times q$ transformation matrix, and can be used to determine the moments of the distribution function $f_{k}^{i}$ and equilibrium distribution function $f_{k}^{i,(eq)}$ in moment space,
\begin{equation}\label{eq3-11}
\mathbf{m}^{i}:=\mathbf{Mf^{i}},\ \ \ \mathbf{m}^{i,(eq)}:=\mathbf{Mf}^{i,(eq)},
\end{equation}
where $\mathbf{f}^{i}=(f_{0}^{i},\; f_{1}^{i},\; \cdots,\; f_{q-1}^{i})^{\top}$ or $(f_{1}^{i},\; f_{2}^{i},\; \cdots,\; f_{q}^{i})^{\top}$, $\mathbf{f}^{i,(eq)}=(f_{0}^{i,(eq)},\; f_{1}^{i,(eq)},\; \cdots, f_{q-1}^{i,(eq)})^{\top}$ or $(f_{1}^{i,(eq)},\; f_{2}^{i,(eq)},\; \cdots, f_{q}^{i,(eq)})^{\top}$ with $\top$ representing the transpose of a matrix. $\mathbf{\Lambda}^{ij}=\textrm{diag}(\lambda_{0}^{ij},\ \lambda_{1}^{ij},\ \cdots,\ \lambda_{q-1}^{ij})$ or $\textrm{diag}(\lambda_{1}^{ij},\ \lambda_{2}^{ij},\ \cdots,\ \lambda_{q}^{ij})$ is a diagonal relaxation matrix, and $\lambda_{k}^{ij}$ is the relaxation parameter corresponding to the $k$th moment of distribution function.

For a specified one-, two- or three-dimensional problem, one can first determine the corresponding lattice model, the transformation matrix $\mathbf{M}$ and the relaxation matrix $\mathbf{\Lambda}^{ij}$ \cite{Cui2016,Chai2018}, then the evolution equation (\ref{eq3-2}) can be implemented with the following two steps,
\begin{subequations}\label{eq3-13}
\begin{equation}
\textrm{Collision:}\;  \mathbf{m}^{i, +}(\mathbf{x} ,\;t) = \mathbf{m}^{i}(\mathbf{x},\;t )-\sum_{j=1}^{n-1}\mathbf{\Lambda}^{ij}[\mathbf{m}^{j}(\mathbf{x},\;t) - \mathbf{m}^{j, (eq)}(\mathbf{x},\;t)],
\end{equation}
\begin{equation}
\textrm{Propagation:}\;  f_{k}^{i}(\mathbf{x} + \mathbf{c}_{k}\delta t,\;t + \delta t) = f_{k}^{i, +}(\mathbf{x} ,\;t), \ \ \ \ f_{k}^{i, +}(\mathbf{x} ,\;t)=\mathbf{M}^{-1}\mathbf{m}^{i, +}(\mathbf{x} ,\;t),
\end{equation}
\end{subequations}
where $f_{k}^{i, +}(\mathbf{x} ,\;t)$ is the post-collision distribution function. We note that although the present model is suitable for one-. two- and three-dimensional multicomponent diffusion problems, for the sake of simplicity, here we only consider the two-dimensional MRT-LB model with $D2Q5$ lattice structure in which the transformation matrix $\mathbf{M}$ and relaxation matrix $\mathbf{\Lambda}^{ij}$ are given by \cite{Yoshida2010,Chai2018}
\begin{subequations}\label{eq3-14}
\begin{equation}
\mathbf{M}=\mathbf{C}_{d}\mathbf{M}_{0}, \; \mathbf{C}_{d}=\textrm{diag}(1, c, c, c^{2}, c^{2}), \ \ \mathbf{M}_{0}=\left(\begin{array}{ccccc}
1 & 1 & 1 & 1 & 1 \\
0 & 1 & 0 & -1 & 0 \\
0 & 0 & 1 & 0 & -1 \\
0 & 1 & -1 & 1 & -1 \\
-4 & 1 & 1 & 1 & 1
\end{array} \right),
\end{equation}
\begin{equation}
\mathbf{\Lambda}^{ij}=\textrm{diag}(\lambda_{0}^{ij},\; \lambda_{1}^{ij},\; \lambda_{1}^{ij},\; \lambda_{2}^{ij},\; \lambda_{2}^{ij}).
\end{equation}
\end{subequations}
Based on Eq.~(\ref{eq3-14}a), we have
\begin{equation}
\mathbf{M}_{0}^{-1}\mathbf{\Lambda}^{ij}\mathbf{M}_{0}=\mathbf{M}^{-1}\mathbf{\Lambda}^{ij}\mathbf{M},
\end{equation}
which can also be used to rewrite the evolution equation (\ref{eq3-2}) in another form
\begin{equation}\label{eq3-1a}
f_{k}^{i}(\mathbf{x} + \mathbf{c}_{k}\delta t,\;t + \delta t) = f_{k}^{i}(\mathbf{x},\;t)-\sum_{j=1}^{n-1}(\mathbf{M}_{0}^{-1}\mathbf{\Lambda}^{ij}\mathbf{M}_{0})_{k\alpha}\big[f_{\alpha}^{j}(\mathbf{x},\;t) - f_{\alpha}^{j, (eq)}(\mathbf{x},\;t)\big].
\end{equation}
We would also like to point out that in Eq.~(\ref{eq3-14}b), the second and third diagonal elements of relaxation matrix $\mathbf{\Lambda}^{ij}$ are denoted by a same parameter $\lambda_{1}^{ij}$ since both of them correspond to the first-order moment of distribution function, while the fourth and fifth relaxation parameters represented by $\lambda_{2}^{ij}$ corresponds to the second-order moment of distribution function. Besides, one can also show that if all the relaxation parameters are equal to each other, the MRT-LB model would reduce to the SRT-LB model \cite{Qian1992}, while if the relaxation parameters corresponding to odd and even-order moments are given by two different values (e.g., $\lambda_{0}^{ij}=\lambda_{2}^{ij}=\lambda_{e}^{ij}$, $\lambda_{1}^{ij}=\lambda_{o}^{ij}$), the MRT-LB model would be the same as the TRT-LB model \cite{Ginzburg2005a}.

In present MRT-LB model, the mole fraction $\xi_{i}$ can be computed through a summation of the distribution function \cite{Chai2016},
\begin{equation}\label{eq3-15}
\xi_{i}(\mathbf{x},\; t)=\sum_{k=0}^{4}f_{k}^{i}(\mathbf{x},\; t),
\end{equation}
the relation between the effective diffusivity $\tilde{D}_{ij}$ and elements of relaxation matrices can be expressed by the following Eq.~(\ref{eq3-26}).

In addition, it should be noted that in the MRT-LB model, besides the relaxation parameter $\lambda_{1}^{ij}$ corresponding to effective diffusivity $\tilde{D}_{ij}$, there are also two free relaxation parameters that need to be determined. In the following simulations, the relaxation parameter $\lambda_{0}^{ij}$ corresponding to the conservation variable is set to be $\lambda_{0}^{ij}=1$ since it almost has no influence on the accuracy and stability of MRT-LB model \cite{Cui2016,Chai2016}. The relaxation parameter $\lambda_{2}^{ij}$ corresponding to the second-order moment, however, is a key parameter \cite{Cui2016}, and to eliminate the discrete effect of half-way bounce-back boundary condition, the following relations are adopted,
\begin{equation}\label{eq3-16}
\lambda_{2}^{ij}=\left\{\begin{array}{c} \lambda_{1}^{ij},\ \ \ \ \ \ \ \  \ \ \ \ \ \ \ \ \ \ \ \ \ \ \ \ \  \ \ \ i\neq j,\\
8(\lambda_{1}^{ij}-2)/(\lambda_{1}^{ij}-8), \ \ i=j.
\end{array}\right.
\end{equation}

Finally, we would also like to point out that compared to the kinetic theory based LB models, the most striking feature of the present LB model is that it can readily handle the multicomponent diffusion problems with different molecular weights, and does not include any complicated interpolations.

\subsection{The Chapman-Enskog analysis}

 We now conduct a detailed Chapman-Enskog analysis, and show how to derive the MS theory based continuum equations from present MRT-LB model. In the Chapman-Enskog analysis, the distribution function $f_{k}^{i}(\mathbf{x},\;t)$, the derivatives of time and space can be expanded as \cite{Chen1998,Succi2001,Guo2013,Kruger2017}
\begin{subequations}\label{eq3-18}
\begin{equation}
f_{k}^{i}=f_{k}^{i,(0)}+\varepsilon f_{k}^{i,(1)}+\varepsilon^{2} f_{k}^{i,(2)}+\cdots,
\end{equation}
\begin{equation}
\partial_t=\varepsilon\partial_{t_{1}}+\varepsilon^{2}\partial_{t_{2}},\ \ \ \nabla=\varepsilon \nabla_{1}=\varepsilon(\partial_{x_{1}},\; \partial_{y_{1}})^{\top},
\end{equation}
\end{subequations}
where $\varepsilon$ is a small parameter.

Taking the Taylor expansion to Eq.~(\ref{eq3-2}), we have
\begin{equation}\label{eq3-19}
D_{k} f_{k}^{i} + \frac{\delta t}{2} D_{k}^{2}f_{k}^{i} = -\sum_{j=1}^{n-1}(\mathbf{M}^{-1}\bar{\mathbf{\Lambda}}^{ij}\mathbf{M})_{k\alpha}\big[f_{\alpha}^{j}- f_{\alpha}^{j,(eq)}\big],
\end{equation}
where $D_{k}=\partial_{t}+ \mathbf{c}_{k}\cdot\nabla$, $\bar{\mathbf{\Lambda}}^{ij}=\mathbf{\Lambda}^{ij}/\delta t$.
Substituting Eq.~(\ref{eq3-18}) into Eq.~(\ref{eq3-19}) yields the following equation,
\begin{eqnarray}\label{eq3-20}
& &\varepsilon D_{k1}f_{k}^{i,(0)}+\varepsilon^{2}\big[\partial_{t_{2}} f_{k}^{i,(0)}+D_{i1}f_{k}^{i,(1)}+\frac{\delta t}{2} D_{k1}^{2}f_{k}^{i,(0)}\big] \\ \nonumber
& = & -\sum_{j=1}^{n-1}(\mathbf{M}^{-1}\bar{\mathbf{\Lambda}}^{ij}\mathbf{M})_{k\alpha}\big[f_{\alpha}^{j,(0)}+\varepsilon f_{\alpha}^{j,(1)}+\varepsilon^{2} f_{\alpha}^{j,(2)} - f_{\alpha}^{j,(eq)}\big]+O(\varepsilon^{3}),
\end{eqnarray}
where $D_{k1}=\partial_{t_{1}}+ \mathbf{c}_{k}\cdot\nabla_{1}$.

From Eq.~(\ref{eq3-20}), one can obtain the zeroth, first and second-order equations in $\varepsilon$,
\begin{subequations}
\begin{equation}\label{ConA1}
\varepsilon^{0}: \ \ \ \sum_{j=1}^{n-1}(\mathbf{M}^{-1}\bar{\mathbf{\Lambda}}^{ij}\mathbf{M})_{k\alpha}\big[f_{\alpha}^{j,(0)}- f_{\alpha}^{j,(eq)}\big]=0,
\end{equation}
\begin{equation}\label{ConA2}
\varepsilon^{1}: \ \ \ \ D_{k1}f_{k}^{i,(0)}=-\sum_{j=1}^{n-1}(\mathbf{M}^{-1}\bar{\mathbf{\Lambda}}^{ij}\mathbf{M})_{k\alpha}f_{\alpha}^{j,(1)},
\end{equation}
\begin{equation}\label{ConA3}
\varepsilon^{2}: \ \ \ \ \partial_{t_{2}} f_{k}^{i,(0)}+D_{k1}f_{k}^{i,(1)}+\frac{\delta t}{2}D_{k1}^{2}f_{k}^{i,(0)}=-\sum_{j=1}^{n-1}(\mathbf{M}^{-1}\bar{\mathbf{\Lambda}}^{ij}\mathbf{M})_{k\alpha}f_{\alpha}^{j,(2)}.
\end{equation}
\end{subequations}
If we introduce the following matrix $\mathbf{\Lambda}$,
\begin{equation}\label{eq3-21}
\mathbf{\Lambda}=\left(\begin{array}{cccc}
\mathbf{\Lambda}^{11} & \mathbf{\Lambda}^{12} & \cdots & \mathbf{\Lambda}^{1(n-1)} \\
\mathbf{\Lambda}^{21} & \mathbf{\Lambda}^{22} & \cdots & \mathbf{\Lambda}^{2(n-1)} \\
\vdots & \vdots  & \vdots & \vdots \\
\mathbf{\Lambda}^{(n-1)1} & \mathbf{\Lambda}^{(n-1)2} & \cdots & \mathbf{\Lambda}^{(n-1)(n-1)}
\end{array} \right),
\end{equation}
and assume that the matrix is non-singular, one can obtain
\begin{equation}\label{ConA1a}
\varepsilon^{0}: \ \ \ f_{k}^{i,(0)} = f_{k}^{i,(eq)}.
\end{equation}

Multiplying the transformation matrix $\mathbf{M}$ on both sides of the zeroth, first and second-order equations in $\varepsilon$, i.e., Eqs.~(\ref{ConA1a}), ~(\ref{ConA2}) and ~(\ref{ConA3}), we have
\begin{subequations}
\begin{equation}\label{ConA1M}
\varepsilon^{0}: \ \ \ \ \mathbf{m}^{i,(0)}=\mathbf{m}^{i,(eq)},
\end{equation}
\begin{equation}\label{ConA2M}
\varepsilon^{1}: \ \ \ \ \mathbf{D}_{1}\mathbf{m}^{i,(0)}=-\sum_{j=1}^{n-1}\bar{\mathbf{\Lambda}}^{ij}\mathbf{m}^{j,(1)},
\end{equation}
\begin{equation}\label{ConA3M}
\varepsilon^{2}: \ \ \ \ \partial_{t_{2}} \mathbf{m}^{i,(0)}+\mathbf{D}_{1}\big(\mathbf{m}^{i,(1)}-\frac{1}{2}\sum_{j=1}^{n-1}\mathbf{\Lambda}^{ij}\mathbf{m}^{j,(1)}\big)=-\sum_{j=1}^{n-1}\bar{\mathbf{\Lambda}}^{ij}\mathbf{m}^{j,(2)},
\end{equation}
\end{subequations}
where Eq.~(\ref{ConA2M}) has been adopted to obtain Eq.~(\ref{ConA3M}). $\mathbf{m}^{i,(k)}=\mathbf{Mf}^{i,(k)}\ (k=0,\; 1,\; 2)$ with $\mathbf{f}^{i,(k)}=\big(f_{0}^{i,(k)},\; \cdots,\; f_{4}^{i,(k)}\big)^{\top}$. Based on Eqs.~(\ref{eq3-3}) and~(\ref{ConA1M}), we can express $\mathbf{m}^{i,(k)} (k=0,\; 1,\; 2)$ as
\begin{equation}\label{m0A}
\mathbf{m}^{i,(0)}=\xi_{i}\big(1,\; 0,\; 0,\; 0,\; -\frac{2}{3}c^{2}\big)^{\top}, \; \mathbf{m}^{i,(1)}=\big(0,\; m_{1}^{i,(1)},\;\ldots,\; m_{4}^{i,(1)}\big)^{\top},\; \mathbf{m}^{i,(2)}=\big(0,\; m_{1}^{i,(2)},\cdots,\; m_{4}^{i,(2)}\big)^{\top}.
\end{equation}
$\mathbf{D}_{1}=\partial_{t}\mathbf{I}+\mathbf{M}\textrm{diag}(c_{0\alpha}\nabla_{0\alpha},\; \cdots,\; c_{4\alpha}\nabla_{4\alpha})\mathbf{M}^{-1}$, and $\mathbf{M}\textrm{diag}(c_{0\alpha}\nabla_{0\alpha},\; \cdots,\; c_{4\alpha}\nabla_{4\alpha})\mathbf{M}^{-1}$ can also be determined explicitly by
\begin{equation}\label{MCM}
\mathbf{M}\textrm{diag}(c_{0\alpha}\nabla_{0\alpha},\; \cdots,\; c_{4\alpha}\nabla_{4\alpha})\mathbf{M}^{-1}=\left(\begin{array}{ccccc}
0 & \partial_{x} & \partial_{y} & 0 & 0  \\
\frac{2c^{2}}{5}\partial_{x} & 0 & 0 & \frac{1}{2}\partial_{x} & \frac{1}{10}\partial_{x} \\
\frac{2c^{2}}{5}\partial_{y} & 0 & 0 & -\frac{1}{2}\partial_{y} & \frac{1}{10}\partial_{y}\\
0 & c^{2}\partial_{x} & -c^{2}\partial_{y} & 0 & 0 \\
0 & c^{2}\partial_{x} & c^{2}\partial_{y} & 0 & 0
\end{array} \right).
\end{equation}
Based on Eq.~(\ref{ConA2M}), we can first derive the first-order equations in $\varepsilon$, but for simplicity, here only the first three ones that used in the following analysis are presented,
\begin{subequations}\label{eq3-22}
\begin{equation}
\partial_{t_{1}}\xi_{i}=0,
\end{equation}
\begin{equation}
\frac{c^{2}}{3}\partial_{x_{1}}\xi_{i}=-\frac{1}{\delta t}\sum_{j=1}^{n-1}\lambda_{1}^{ij}m_{1}^{j,(1)},
\end{equation}
\begin{equation}
\frac{c^{2}}{3}\partial_{y_{1}}\xi_{i}=-\frac{1}{\delta t}\sum_{j=1}^{n-1}\lambda_{1}^{ij}m_{2}^{j,(1)}.
\end{equation}
\end{subequations}
where Eqs.~(\ref{m0A}) and (\ref{MCM}) have bee used. Similarly, from Eq.~(\ref{ConA3M}), one can also derive the second-order equations in $\varepsilon$, but here we only present the first one corresponding to the conservative variable $\phi$,
\begin{equation}\label{RA2}
\partial_{t_{2}}\xi_{i}+\partial_{x_{1}}\big(m_{1}^{i,(1)}-\frac{1}{2}\sum_{j=1}^{n-1}\lambda_{1}^{ij}m_{1}^{j,(1)}\big)+\partial_{y_{1}}\big(m_{2}^{i,(1)}-\frac{1}{2}\sum_{j=1}^{n-1}\lambda_{1}^{ij}m_{2}^{j,(1)}\big)=0,
\end{equation}
which can also be written in the matrix form,
\begin{equation}\label{RA2a}
\partial_{t_{2}}\mathbf{\xi}+\partial_{x_{1}}\big(\mathbf{I}-\frac{1}{2}\mathbf{\Lambda}_{1}\big)\mathbf{m}_{1}^{(1)}+\partial_{y_{1}}\big(\mathbf{I}-\frac{1}{2}\mathbf{\Lambda}_{1}\big)\mathbf{m}_{2}^{(1)}=0,
\end{equation}
where $\mathbf{\xi}$, $\mathbf{\Lambda}_{1}$ and $\mathbf{m}_{k}^{(1)}\; (k=1,\; 2)$ are defined by
\begin{subequations}\label{eq3-23}
\begin{equation}
\mathbf{\xi}=(\xi_{1},\; \cdots,\; \xi_{n-1})^{\top},
\end{equation}
\begin{equation}
\mathbf{m}_{k}^{(1)}=\big(m_{k}^{1, (1)},\; \cdots,\; m_{k}^{n-1, (1)}\big)^{\top},
\end{equation}
\begin{equation}
\mathbf{\Lambda}_{1}=\left(\begin{array}{cccc}
\mathbf{\lambda}_{1}^{11} & \mathbf{\lambda}_{1}^{12} & \cdots & \mathbf{\lambda}_{1}^{1(n-1)} \\
\mathbf{\lambda}_{1}^{21} & \mathbf{\lambda}_{1}^{22} & \cdots & \mathbf{\lambda}_{1}^{2(n-1)} \\
\vdots & \vdots  & \vdots & \vdots \\
\mathbf{\lambda}_{1}^{(n-1)1} & \mathbf{\lambda}_{1}^{(n-1)2} & \cdots & \mathbf{\lambda}_{1}^{(n-1)(n-1)}
\end{array} \right).
\end{equation}
\end{subequations}
On the other hand, from Eqs.~(\ref{eq3-22}b) and~(\ref{eq3-22}c), one can also obtain $\mathbf{m}_{1}^{(1)}$ and $\mathbf{m}_{2}^{(1)}$,
\begin{equation}\label{eq3-24}
\mathbf{m}_{1}^{(1)}=-\frac{c^{2}}{3}\delta t\mathbf{\Lambda}_{1}^{-1}\partial_{x_{1}}\mathbf{\xi},
\end{equation}
\begin{equation}\label{eq3-25}
\mathbf{m}_{2}^{(1)}=-\frac{c^{2}}{3}\delta t\mathbf{\Lambda}_{1}^{-1}\partial_{y_{1}}\mathbf{\xi},
\end{equation}
where $\mathbf{\Lambda}_{1}$ has been assumed to be invertible. Substituting Eqs.~(\ref{eq3-24}) and~(\ref{eq3-25}) into Eq.~(\ref{RA2a}) leads to the following result,
\begin{equation}\label{RA2aa}
\partial_{t_{2}}\mathbf{\xi}=\partial_{x_{1}}\big[\frac{c^{2}}{3}\delta t\big(\mathbf{\Lambda}_{1}^{-1}-\frac{1}{2}\mathbf{I}\big)\partial_{x_{1}}\mathbf{\xi}\big]+\partial_{y_{1}}\big[\frac{c^{2}}{3}\delta t\big(\mathbf{\Lambda}_{1}^{-1}-\frac{1}{2}\mathbf{I}\big)\partial_{y_{1}}\mathbf{\xi}\big]=0,
\end{equation}
from which one can also obtain the equation for species $i$,
\begin{equation}\label{RA3}
\partial_{t_{2}}\xi_{i}=\nabla_{1}\cdot\big(\sum_{j=1}^{n-1}\tilde{D}_{ij}\nabla_{1}\xi_{j}\big),
\end{equation}
where the diffusivity $\tilde{D}_{ij}$ or the matrix $\tilde{\mathbf{D}}$ are given by
\begin{equation}\label{eq3-26}
\tilde{D}_{ij}=\frac{c^{2}}{3}\delta t\big(\mathbf{\Lambda}_{1}^{-1}-\frac{1}{2}\mathbf{I}\big)_{ij},\ \ \ \tilde{\mathbf{D}}=\frac{c^{2}}{3}\delta t\big(\mathbf{\Lambda}_{1}^{-1}-\frac{1}{2}\mathbf{I}\big).
\end{equation}
Through a combination of Eqs.~(\ref{eq3-22}a) and (\ref{RA3}), i.e., $\varepsilon\times(\ref{eq3-22}a)+\varepsilon^{2}\times(\ref{RA3})$, we can correctly recover the MS theory based continuum equations for all species [see (\ref{eq2-13})].

Now let us focus on how to calculate the gradient term $\nabla\xi_{i}\ (1\leq i\leq n-1)$, which can be used to determine the diffusion flux [see Eq.~(\ref{eq2-11})]. Generally speaking, there are two possible ways that can be applied to obtain $\nabla\xi_{i}$. The first one is to directly use the nonlocal finite-difference scheme to compute $\nabla\xi_{i}$, while the second is, in the framework of LB method, to derive $\nabla\xi_{i}$ locally through the nonequilibrium part of the distribution function \cite{Chai2013, Chai2014}. Actually, the second approach has also been adopted to improve stability of LB method \cite{Yang2014} and to predict effective diffusivity of porous media \cite{Chai2016b}.

Here we only consider the latter one for its locality in the computation of gradient term. If we multiply $\varepsilon$ on both sides of Eqs.~(\ref{eq3-22}b) and (\ref{eq3-22}c), one can obtain
\begin{subequations}\label{3-27}
\begin{equation}
\frac{c^{2}}{3}\partial_{x}\xi_{i}=-\frac{1}{\delta t}\sum_{j=1}^{n-1}\lambda_{1}^{ij}\big[m_{1}^{j}-m_{1}^{j,(0)}\big]=-\frac{1}{\delta t}\sum_{j=1}^{n-1}\lambda_{1}^{ij}m_{1}^{j}=-\frac{1}{\delta t}\sum_{j=1}^{n-1}\lambda_{1}^{ij}\sum_{k=0}^{4}\mathbf{c}_{k,x}f_{k}^{j},
\end{equation}
\begin{equation}
\frac{c^{2}}{3}\partial_{y}\xi_{i}=-\frac{1}{\delta t}\sum_{j=1}^{n-1}\lambda_{1}^{ij}\big[m_{2}^{j}-m_{2}^{j,(0)}\big]=-\frac{1}{\delta t}\sum_{j=1}^{n-1}\lambda_{1}^{ij}m_{2}^{j}=-\frac{1}{\delta t}\sum_{j=1}^{n-1}\lambda_{1}^{ij}\sum_{k=0}^{4}\mathbf{c}_{k,y}f_{k}^{j},
\end{equation}
\end{subequations}
where the assumptions of $\varepsilon m_{1}^{j, (1)}=m_{1}^{j}-m_{1}^{j,(0)}$ and $\varepsilon m_{2}^{j, (1)}=m_{2}^{j}-m_{2}^{j,(eq)}$ \cite{Chai2013,Chai2014,Chai2016}, and the fact $m_{1}^{j,(0)}=m_{2}^{j,(0)}=0$ have been adopted. From above equations, we can finally determine $\nabla\xi_{i}=(\partial_{x}\xi_{i}, \; \partial_{y}\xi_{i})^{\top}$,
\begin{equation}\label{xix}
\nabla\xi_{i}=-\frac{3}{\delta t c^{2}}\sum_{j=1}^{n-1}\lambda_{1}^{ij}\sum_{k=0}^{4}\mathbf{c}_{k}f_{k}^{j},
\end{equation}
which can be further used to calculate the diffusion flux $\mathbf{J}$ [see Eq.~(\ref{eq2-11})].

\section{Numerical validation and discussion}

In this section, several benchmark problems appeared in the previous works \cite{Zheng2010,Crank1975,Taylor1993,Arnold1967,Geiser2015,Boudin2012} are used to validate present MRT-LB model. In our simulations, the half-way anti-bounce-back scheme is used for the Dirichlet boundary condition \cite{Zhang2012,Ginzburg2005b,Chen2013},
\begin{equation}
f_{k}^{i}(\mathbf{x}_{f},\;t + \delta t) = -f_{\bar{k}}^{i, +}(\mathbf{x}_{f} ,\;t)+2\omega_{\bar{k}}\xi_{i,b},
\end{equation}
while for no-flux boundary condition, the standard half-way bounce-back scheme is adopted \cite{Ginzburg2005b,Yoshida2010},
\begin{equation}
f_{k}^{i}(\mathbf{x}_{f},\;t + \delta t) = f_{\bar{k}}^{i, +}(\mathbf{x}_{f} ,\;t),
\end{equation}
where $f_{k}^{i}(\mathbf{x}_{f},\;t + \delta t)$ is the unknown distribution function at the boundary node
$\mathbf{x}_{f}$. $\xi_{i,b}$ is the mole fraction at the boundary, and is specified by the Dirichlet boundary condition, $\bar{k}$ is the opposite
direction of $k$. In the initialization process, the distribution function is given by its equilibrium part,
\begin{equation}
f_{k}^{i}(\mathbf{x},\;t)|_{t=0} = f_{k}^{i, (eq)}(\mathbf{x},\;t)|_{t=0}=\omega_{k}\xi_{i}|_{t=0}.
\end{equation}

Additionally, to quantitatively measure the deviation between the numerical and analytical solutions, the relative error based on $L^{2}$ norm is used here,
\begin{equation}
E(\phi)=\sqrt{\frac{\sum_{(x,\; y)}|\phi_{a}(x,\; y,\; t)-\phi_{n}(x,\; y,\; t)|^{2}}{\sum_{(x,\; y)}|\phi_{a}(x,\; y,\; t)|^{2}}},
\end{equation}
where $\phi_{a}$ and $\phi_{n}$ denote the analytical and numerical results of the variable $\phi$.

\subsection{A simple two-component diffusion problem}

We first consider a simple two-component diffusion problem with a constant diffusivity $\mathcal{D}$ [see Eq.~(\ref{eq2-14})] \cite{Crank1975}, which is also used to test the kinetic theory based LB models \cite{Zheng2010,Arcidiacono2006}. For this problem, the MS theory based continuum equation would reduce to the Fick's law based conservation equation. For simplicity, here we only consider the mole fraction $\xi_{1}$ since the mole fraction $\xi_{2}$ can be obtained by $\xi_{2}=1-\xi_{1}$. Under the following initial and boundary conditions,
\begin{subequations}\label{eq4-1}
\begin{equation}
t=0: \; \xi_{1}=C_{0},\; x<0, \;  \xi_{1}=C_{1},\; x\geq0,
\end{equation}
\begin{equation}
x=-\infty,\; \xi_{1}=C_{0},\;  \;  x=+\infty,\; \xi_{1}=C_{1},
\end{equation}
\end{subequations}
one can obtain the analytical solutions of $\xi_{1}$ and diffusion flux $\mathbf{J}_{1}$,
\begin{subequations}\label{eq4-2}
\begin{equation}
\xi_{1}=\frac{C_{0}+C_{1}}{2}+\frac{C_{1}-C_{0}}{2}\textrm{erf}\big(\frac{x}{2\sqrt{\mathcal{D} t}}\big),
\end{equation}
\begin{equation}
\mathbf{J}_{1}=-D\nabla\xi_{1}=-\frac{C_{1}-C_{0}}{2}\sqrt{\frac{\mathcal{D}}{\pi t}}e^{-\frac{x^{2}}{4\mathcal{D}t}},
\end{equation}
\end{subequations}
where erf is the error function, and is defined by
\begin{equation}
\textrm{erf}(y)=\frac{2}{\sqrt{\pi}}\int_{0}^{y}e^{-\eta^{2}}d\eta.
\end{equation}

In the following, we consider the problem with the diffusivity $\mathcal{D}=0.05$, $C_{0}=0.9$, $C_{1}=0.1$, and adopt the periodic boundary condition in $y$ direction. The computational domain is fixed to be $[-6,\; 6]\times[-1,\; 1]$, and to ensure our simulations to be consistent with the physical problem, $x/2\sqrt{\mathcal{D} t}$ should be large enough. We first performed some simulations with the lattice size $240\times40$ and the relaxation parameter $\lambda_{1}^{11}=1.25$, and presented the results at different time in Figs. 1 and 2. As seen from these two figures, the numerical results of mole fraction $\xi_{1}$ and diffusion flux $\mathbf{J}_{1}$ are in good agreement with the corresponding analytical solutions.

\begin{figure}
\includegraphics[width=3.5in]{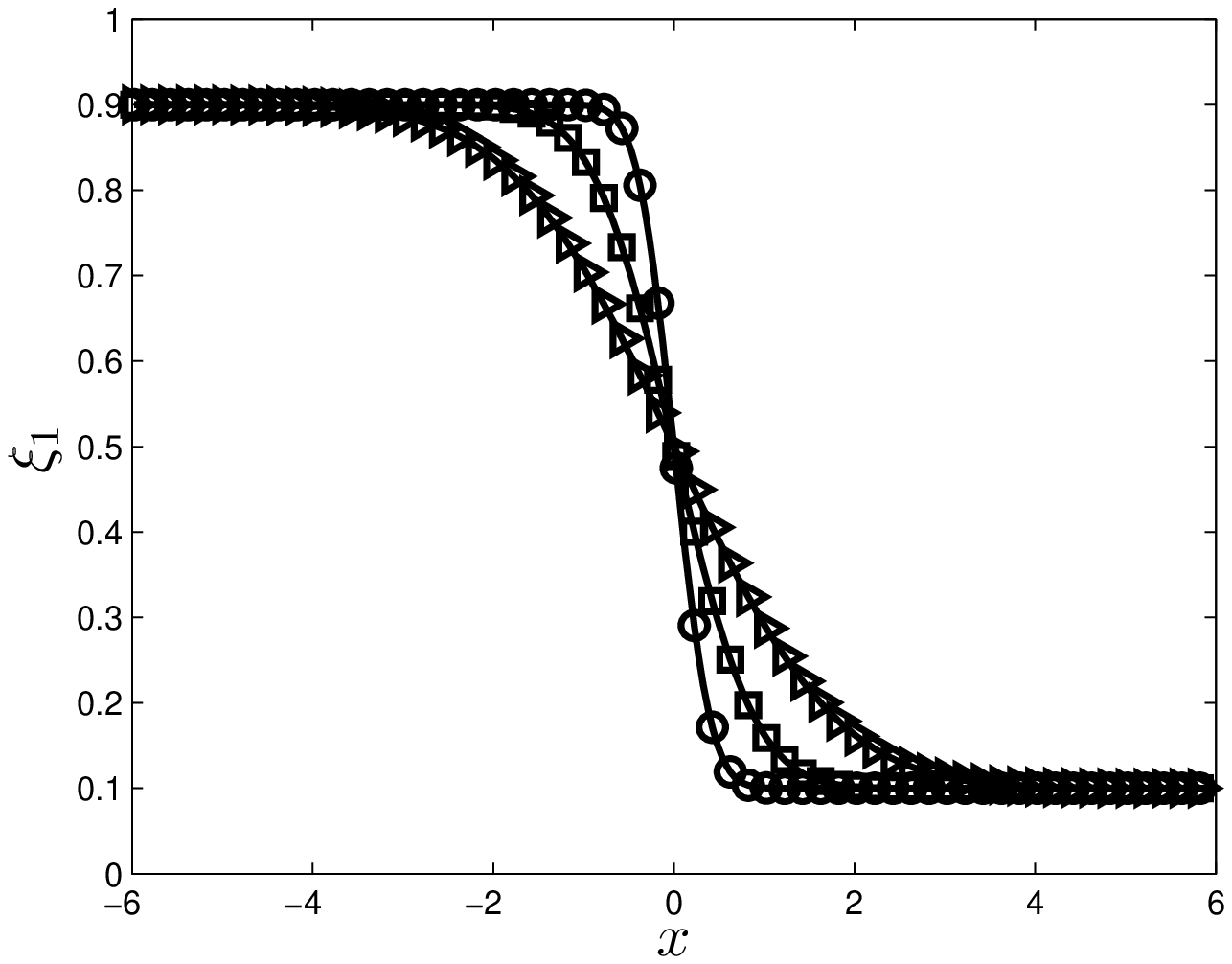}
\centering\caption{\label{fig:1} The profiles of mole fraction $\xi_{1}$ at different time [Solid line: Eq.~(\ref{eq4-2}a), $\bigcirc$: $t=1.0$, $\Box$: $t=5.0$, $\rhd$: $t=20.0$].}
\end{figure}

\begin{figure}
\includegraphics[width=3.5in]{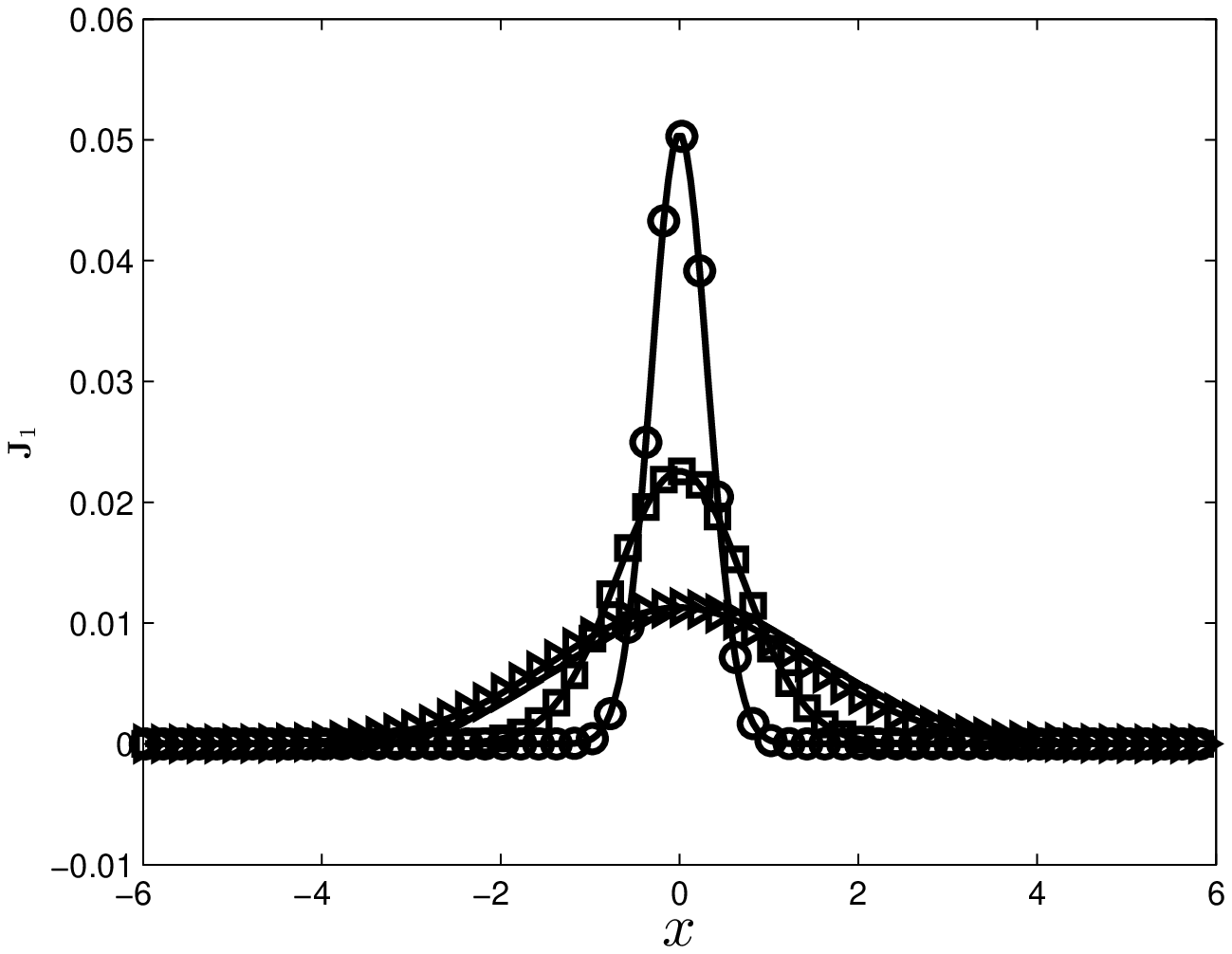}
\centering\caption{\label{fig:2} The profiles of diffusion flux $\xi_{1}$ at different time [Solid line: Eq.~(\ref{eq4-2}b), $\bigcirc$: $t=1.0$, $\Box$: $t=5.0$, $\rhd$: $t=20.0$].}
\end{figure}

Then we also conducted several simulations with the SRT-LB model in which $\lambda_{0}^{11}=\lambda_{1}^{11}=\lambda_{2}^{11}=1.25$, and presented a comparison between these two LB models in Table I. From this table, one can find that the results of MRT-LB model is more accurate than the SRT-LB model, which is mainly caused by the adoption of Eq.~(\ref{eq3-16}) in the MRT-LB model.

 \begin{table*}
\caption{A comparison between the errors of MRT-LB and SRT-LB models for mole fraction $\xi_{1}$ and flux $\mathbf{J}_{1}$.}
\centering
\begin{tabular}{c|cc|cc}
\hline
 \multirow{2}{*}{Time}  & \multicolumn{2}{c}{MRT-LB model} & \multicolumn{2}{c}{SRT-LB model}\\
 \cline{2-5}
  & $E(\xi_{1})$ & $E(\mathbf{J}_{1})$ & $E(\xi_{1})$ & $E(\mathbf{J}_{1})$ \\ \hline
 $t=1$ & $1.4986\times10^{-5}$ & $2.7702\times10^{-4}$ & $1.1792\times10^{-4}$ & $1.0622\times10^{-3}$\\
 $t=5$ & $4.5260\times10^{-6}$ & $5.5143\times10^{-5}$ & $3.5758\times10^{-5}$ & $2.1227\times10^{-4}$\\
 $t=20$ & $2.8366\times10^{-6}$ & $4.6922\times10^{-5}$ & $1.3219\times10^{-5}$ & $6.7868\times10^{-5}$\\ \hline

\end{tabular}
\end{table*}

This problem is also adopted to test convergence rate of the MRT-LB model. To this end, we carried out some simulations under different lattice sizes ($\delta x=1/10,\; 1/20,\; 1/30,\; 1/40$), and the errors of mole fraction $\xi_{1}$ and diffusion flux $\mathbf{J}_{1}$ are shown in Figs. 3 and 4 where the effects of relaxation parameter $\lambda_{1}^{11}$ are also considered. As shown in these two figures, the MRT-LB model has a second-order convergence rate in space, both for mole fraction $\xi_{1}$ and diffusion flux $\mathbf{J}_{1}$. Besides, it is also found that although the relaxation parameter $\lambda_{1}^{11}$ has some influences on the results of mole fraction $\xi_{1}$ and diffusion flux $\mathbf{J}_{1}$, it does not affect the second-order accuracy of MRT-LB model.

\begin{figure}
\includegraphics[width=3.5in]{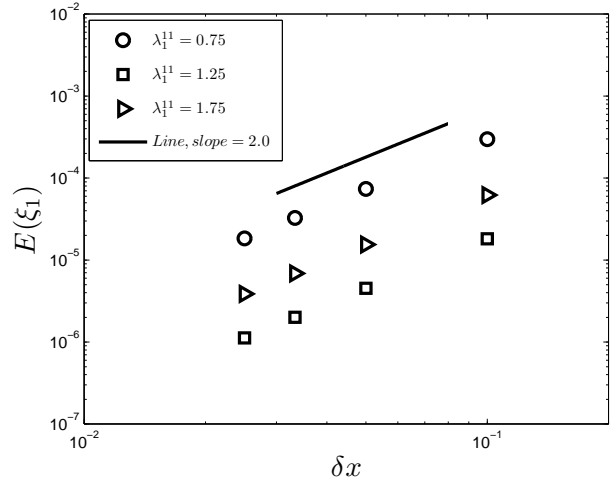}
\centering\caption{\label{fig:3} The errors of MRT-LB model for mole fraction $\xi_{1}$ at the time $t=5$, the slope of the inserted line is 2.0, which indicates that the MRT-LB model has a second-order convergence rate in space }
\end{figure}

\begin{figure}
\includegraphics[width=3.5in]{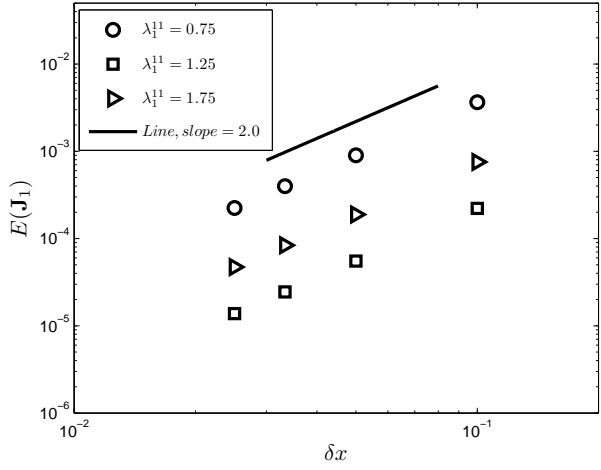}
\centering\caption{\label{fig:4} The errors of MRT-LB model for mole fraction $\mathbf{J}_{1}$ at the time t=5, the slope of the inserted line is 2.0, which indicates that the MRT-LB model has a second-order convergence rate in space.}
\end{figure}

\subsection{A three-component coupling diffusion problem}

We continue to consider a three-component coupling diffusion problem \cite{Geiser2015,Boudin2012} that is a close approximation to the classic experiment conducted by Duncan and Toor \cite{Duncan1962}. In the study of this problem, the diffusivities are set to be $D_{12}=D_{13}=0.833$ and $D_{23}=0.168$, the physical domain of the problem is $[0,\; 1]\times[0,\; 1]$, and the initial and boundary conditions of $\xi_{1}$ and $\xi_{2}$ are given by
\begin{subequations}
\begin{equation}
\xi_{1}=\left\{\begin{array}{c} 0.8,\ \ \ \ \ \ \ \ \  \ \ \ \ 0\leq x <0.25,\\
1.6(0.75-x), \ \ 0.25\leq x <0.75,\\
0, \ \ \ \ \ \ \ \ \ \ \ \ \ \  \ \ 0.75\leq x \leq 1,
\end{array}\right.
\end{equation}
\begin{equation}
\xi_{2}=0.2, \ \ \ \ \ \ \ \ 0\leq x \leq 1,
\end{equation}
\end{subequations}
\begin{subequations}
\begin{equation}
x=0, \ \ \mathbf{J}_{1}=\mathbf{J}_{2}=0,\ \ x=1, \ \  \mathbf{J}_{1}=\mathbf{J}_{2}=0.
\end{equation}
\begin{equation}
\xi_{i}|_{x=0}=\xi_{i}|_{x=1} \ (i=1,\ 2,\ 3).
\end{equation}
\end{subequations}
Under the condition of $D_{12}=D_{13}$, one can rewrite Eq.~(\ref{eq2-9}) as
\begin{equation}\label{SD}
\tilde{\mathbf{D}}=\left(\begin{array}{cc}
D_{12} & 0 \\
\beta\xi_{2}\big(1-\frac{D_{12}}{D_{23}}\big) & \beta
\end{array} \right),
\end{equation}
where the parameter $\beta$ is defined by
\begin{equation}
\beta=\big[\frac{1}{D_{23}}+\xi_{1}\big(\frac{1}{D_{12}}-\frac{1}{D_{23}}\big)\big]^{-1}.
\end{equation}

We carried out some simulations with the lattice size $200\times200$, and presented some results in Figs. 5-9. From the Figs. 5-7, one can first observe that the numerical results at $x=0.72$ are very close to those reported in some previous works \cite{Boudin2012,Geiser2015}. Then let us focus on the changes of mole fraction, diffusion flux and the negative of mole fraction gradient in time. As shown in Fig. 5, the mole fraction $\xi_{1}$ normally increases in time, and finally approaches to the equilibrium value $\xi_{1}^{*}=0.4$ [also see Fig. 8(a) where the time $t$ is increased to $t=5$]. Besides, from Fig. 6, we can also observe that the diffusion flux $\mathbf{J}_{1}$ and the negative of mole fraction gradient $-\partial_{x}\xi_{1}$ decrease with the increase of time, and becomes zero when $t$ is large enough [see Figs. 8(b) and 8(c)]. These results on species 1 are consistent with the theory based on Fick's law since there are no cross effects induced by other species ($D_{12}=D_{13}$). However, from Fig. 5, one can also find some curious results on the mole fraction $\xi_{2}$ which are caused by the cross effects from other species, and can also be seen clearly from Eq.~(\ref{eq2-12}) under the specified matrix $\mathbf{D}$ given by Eq.~(\ref{SD}).
Initially, the mole fraction $\xi_{2}$ is at its equilibrium value $\xi_{2}^{*}=0.2$, based on the Fick's law, there should be no mass diffusion for species 2. While the mole fraction $\xi_{2}$ first decreases, and reaches to the minimum value $\xi_{2}^{min}=0.1581$ at about $t=0.2$. After that, the mole fraction $\xi_{2}$ begins to increase, and is up to the equilibrium value $\xi_{2}^{*}=0.2$ when the time is large enough [see Fig. 9(a)]. We note that these interesting results are similar to the classic experimental results reported in Ref. \cite{Duncan1962}, and cannot depicted by the theory based on the simple Fick's law. To elucidate these phenomena more clearly, we also plotted the variations of diffusion flux and negative of the mole fraction gradient in time in Fig. 7, from which one can find that when the time is located in the range of $0<t<t_{1}$ or $t_{2}<t<t_{3}$, the so-called reverse (or uphill) diffusion ($-\mathbf{J}_{2}\times\partial_{x}\xi_{2}<0$) is observed, while at the time $t=t_{1}$ or $t_{2}$, the osmotic diffusion ($\partial_{x}\xi_{2}=0,\ \mathbf{J}_{2}\neq 0$) can be observed, and at the time $t=t_{3}$, one can observe the phenomenon of diffusion barrier ($\partial_{x}\xi_{2}\neq0,\ \mathbf{J}_{2}=0$). We refer the reader to Ref. \cite{Krishna1997} for more physical explanations on these curious phenomena.

Finally, we would also like to point out that when the time $t$ is large enough, the problem would reach a steady state, and simultaneously, the diffusion flux and mole fraction gradient would become zero [see Figs. 8(b), 8(c), 9(b) and 9(c)].

\begin{figure}
\includegraphics[width=3.5in]{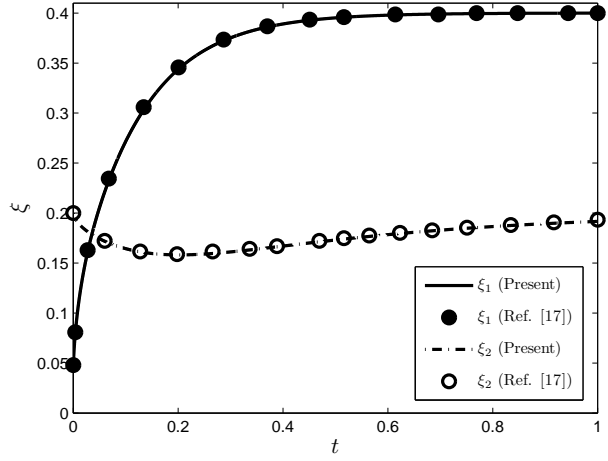}
\centering\caption{\label{fig:5} The mole fraction $\xi_{i}\ (i=1,\ 2)$ at different time ($x=0.72$).}
\end{figure}

\begin{figure}
\includegraphics[width=2.5in]{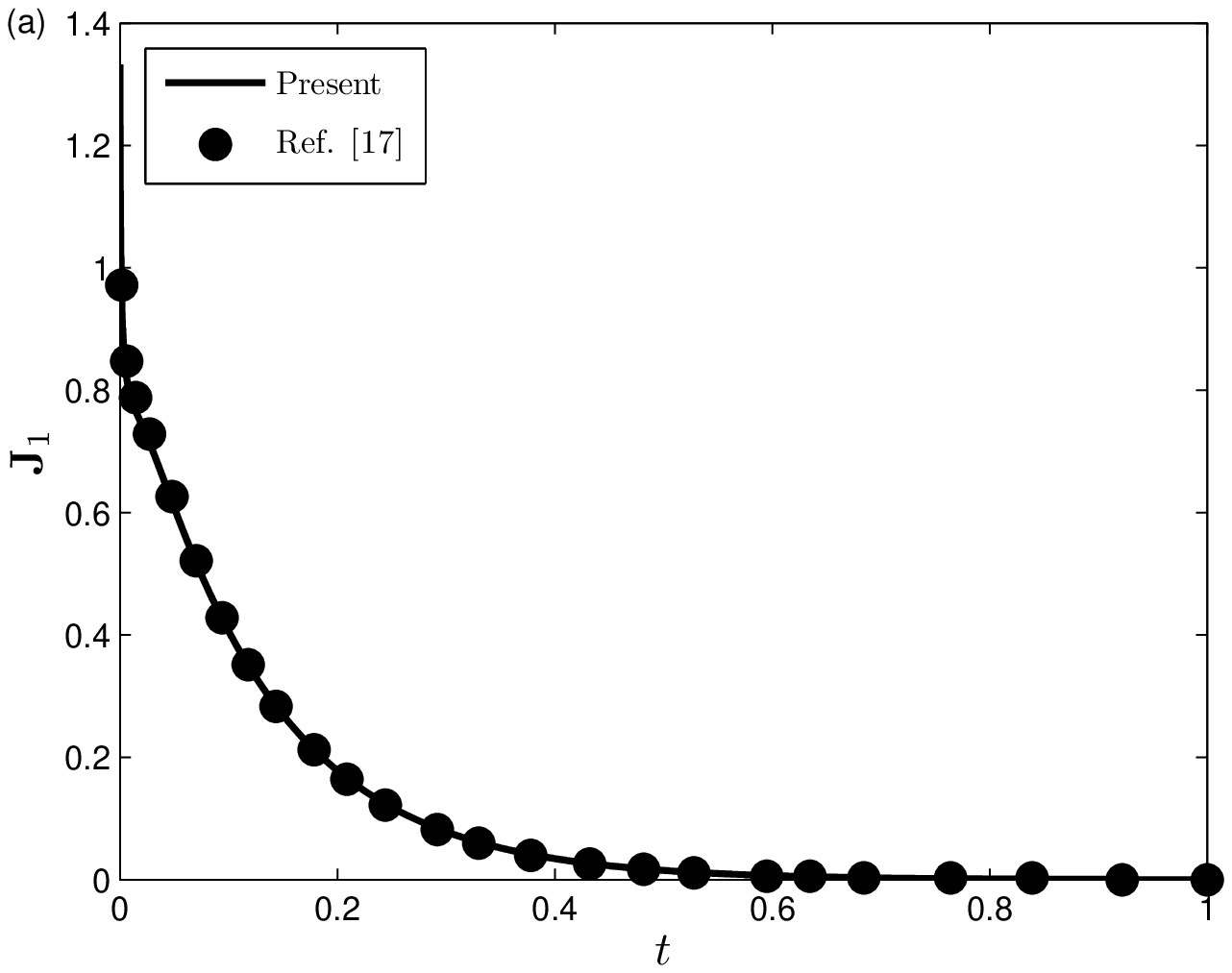}
\includegraphics[width=2.5in]{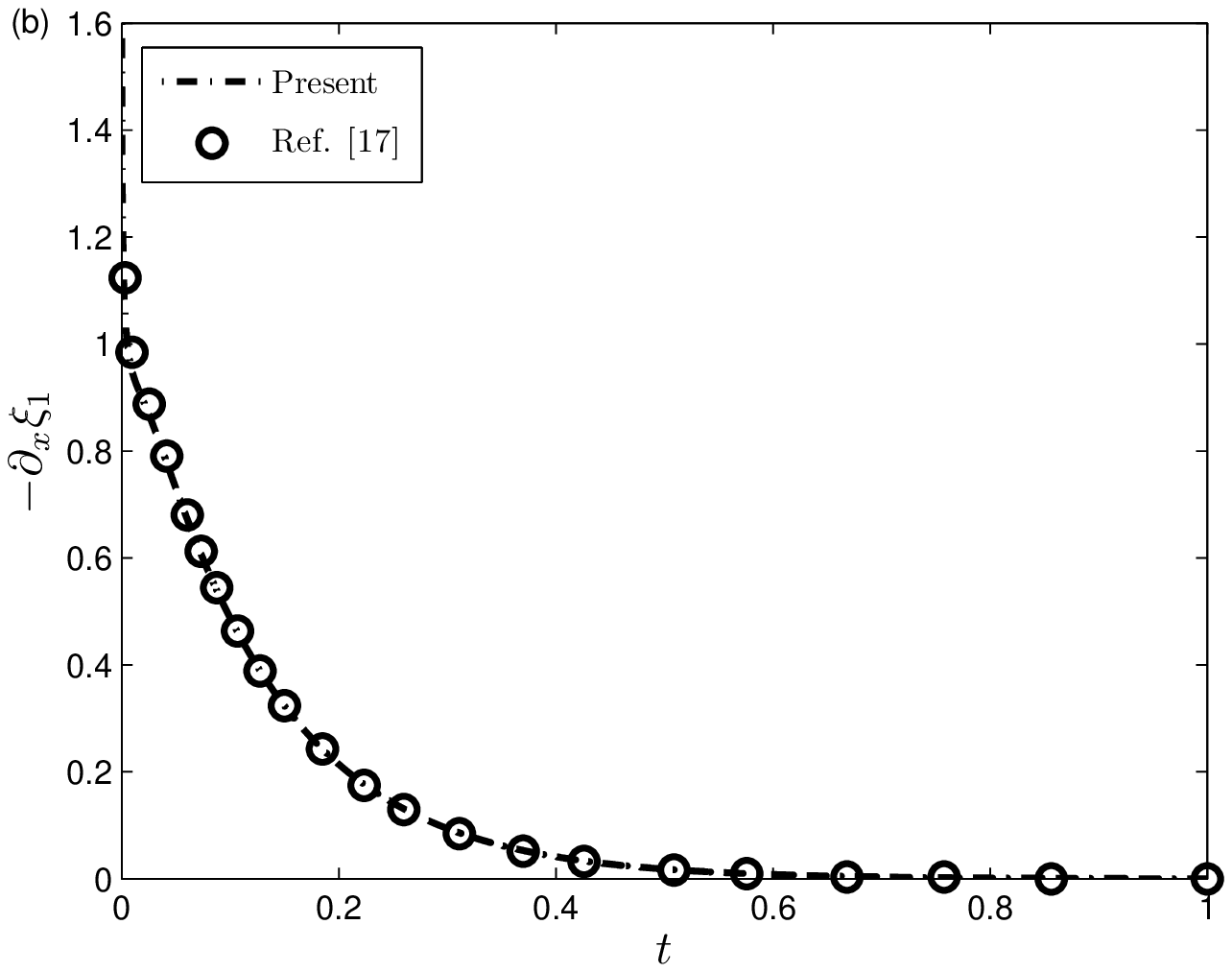}
\centering\caption{\label{fig:6} The diffusion flux $\mathbf{J}_{1}$ and the negative of mole fraction gradient ($-\partial_{x}\xi_{1}$) at different time ($x=0.72$).}
\end{figure}

\begin{figure}
\includegraphics[width=3.5in]{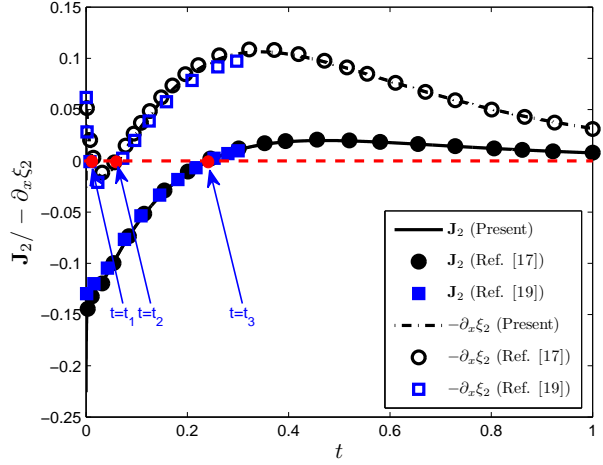}
\centering\caption{\label{fig:7} The diffusion flux $\mathbf{J}_{2}$ and the negative of mole fraction gradient ($-\partial_{x}\xi_{2}$) at different time ($x=0.72$).}
\end{figure}

\begin{figure}
\includegraphics[width=3.5in]{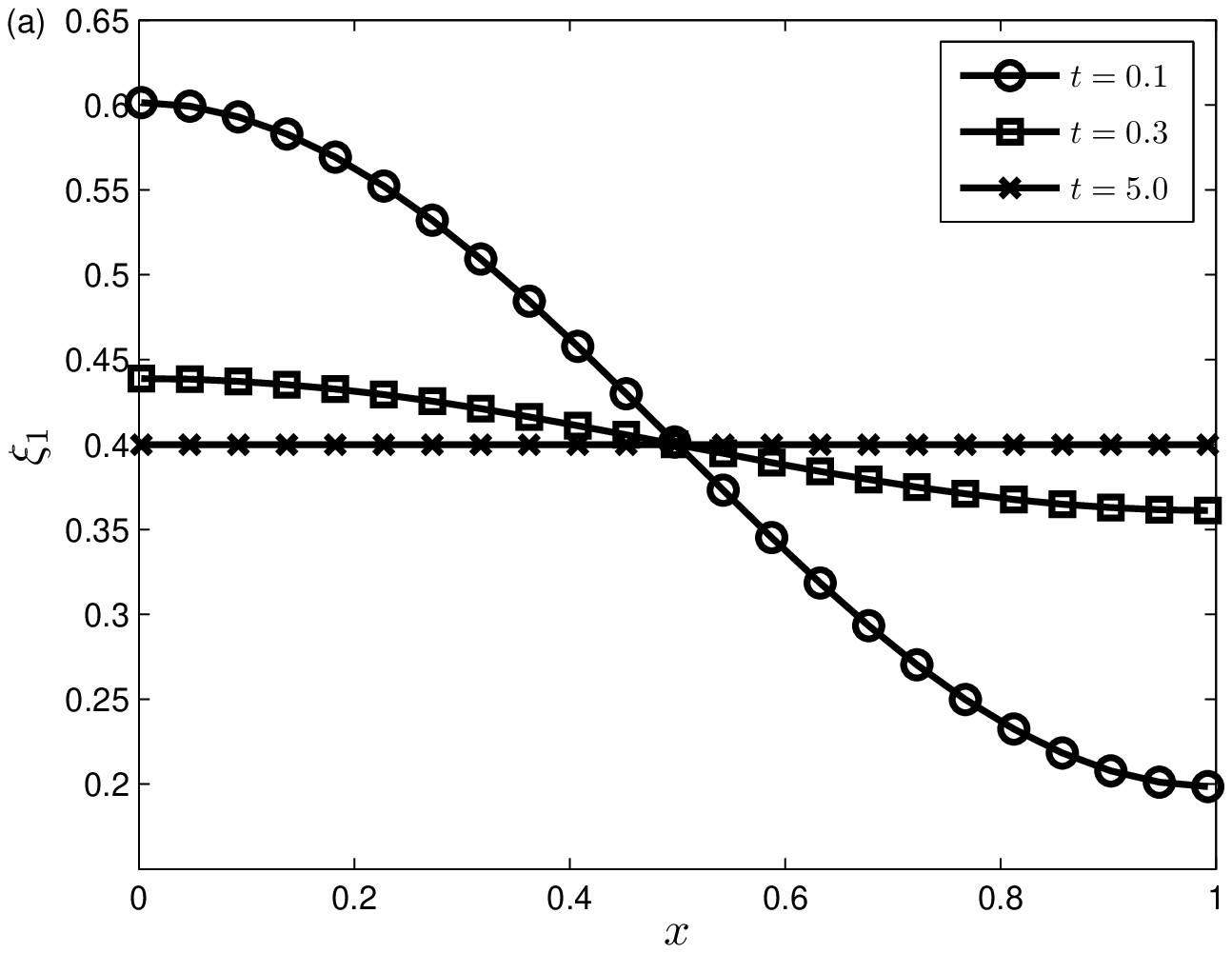}
\includegraphics[width=3.5in]{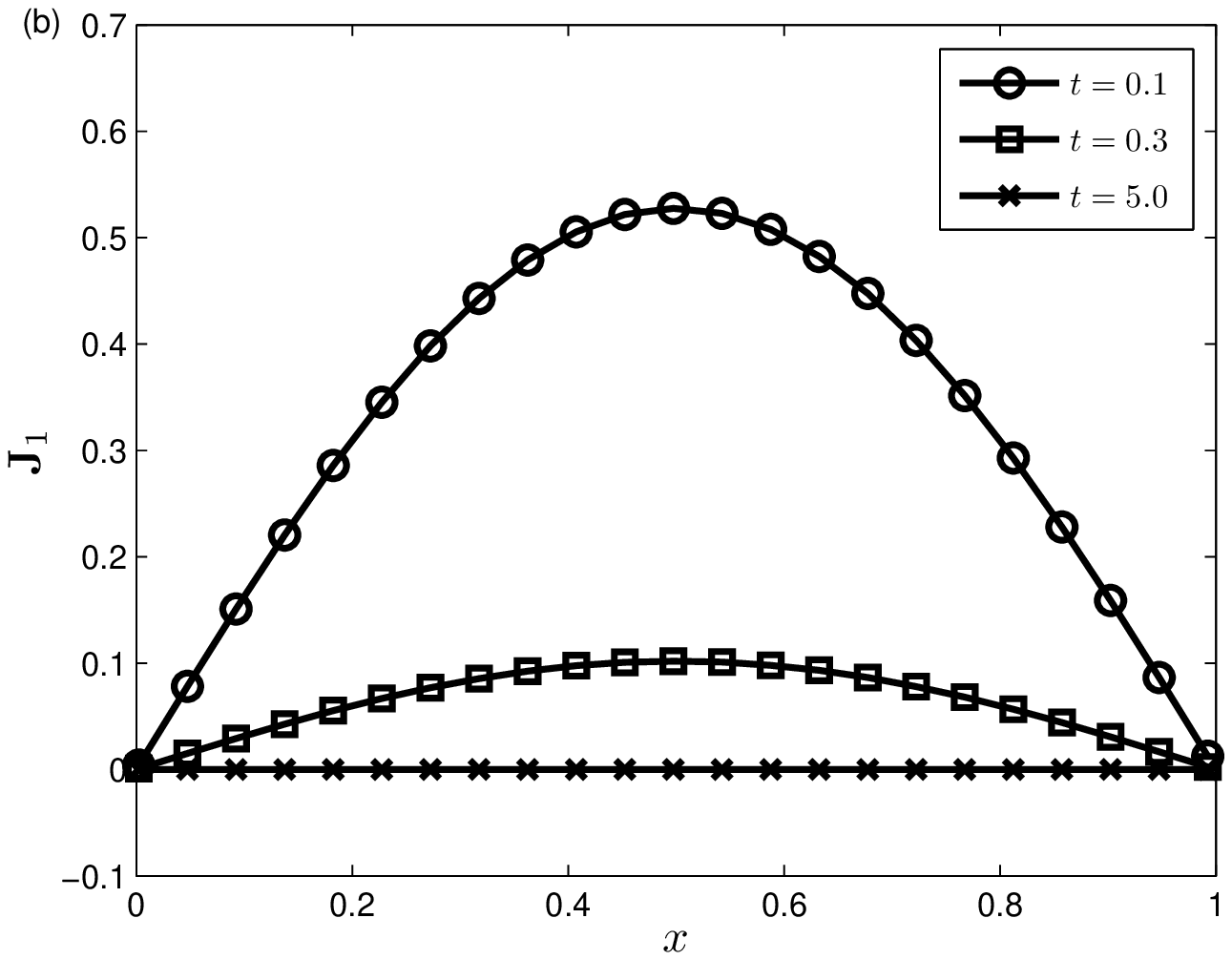}
\includegraphics[width=3.5in]{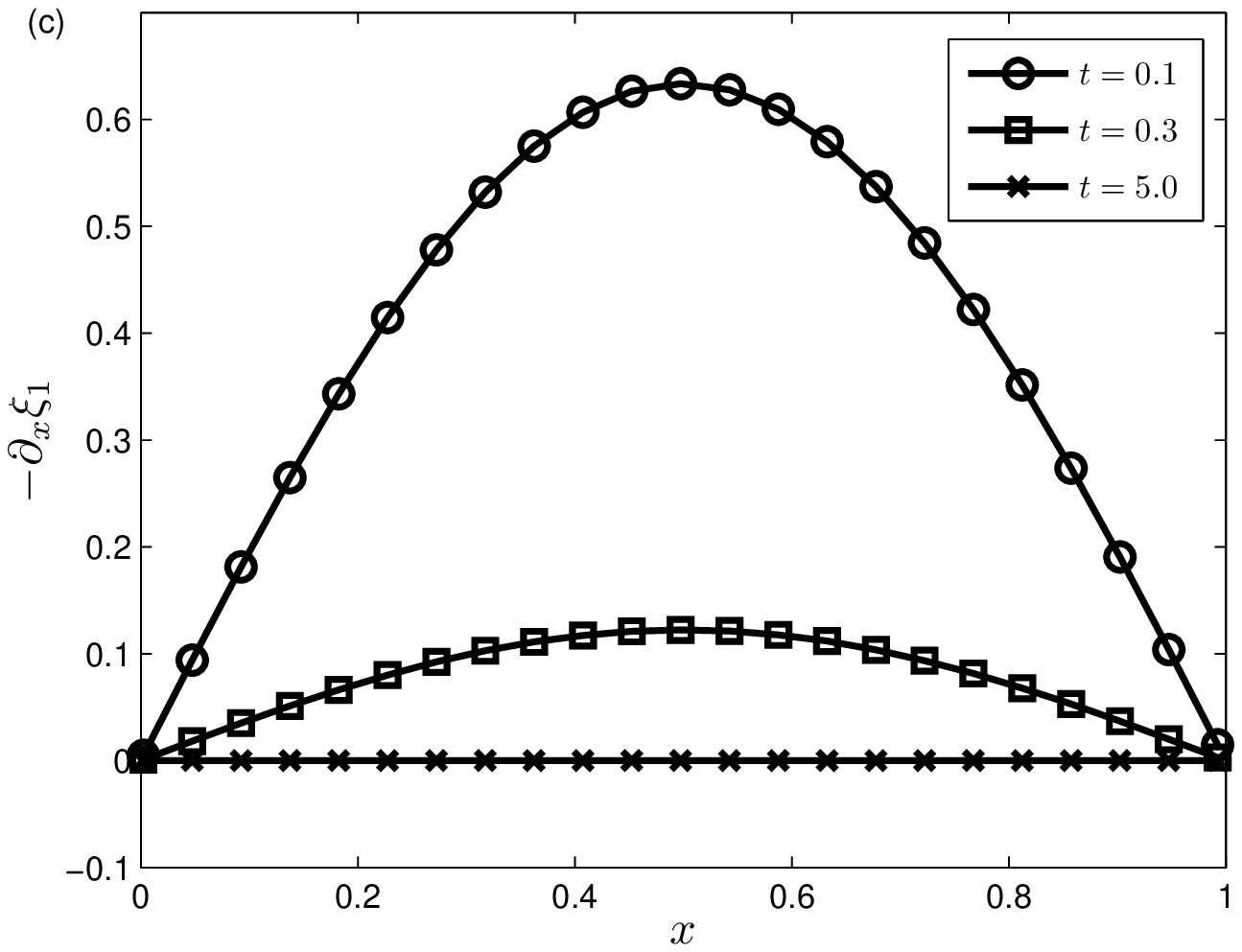}
\centering\caption{\label{fig:8} The distributions of mole fraction $\xi_{1}$, diffusion flux $\mathbf{J}_{1}$ and negative of mole fraction gradient $-\partial_{x}\xi_{1}$ at different time.}
\end{figure}

\begin{figure}
\includegraphics[width=3.5in]{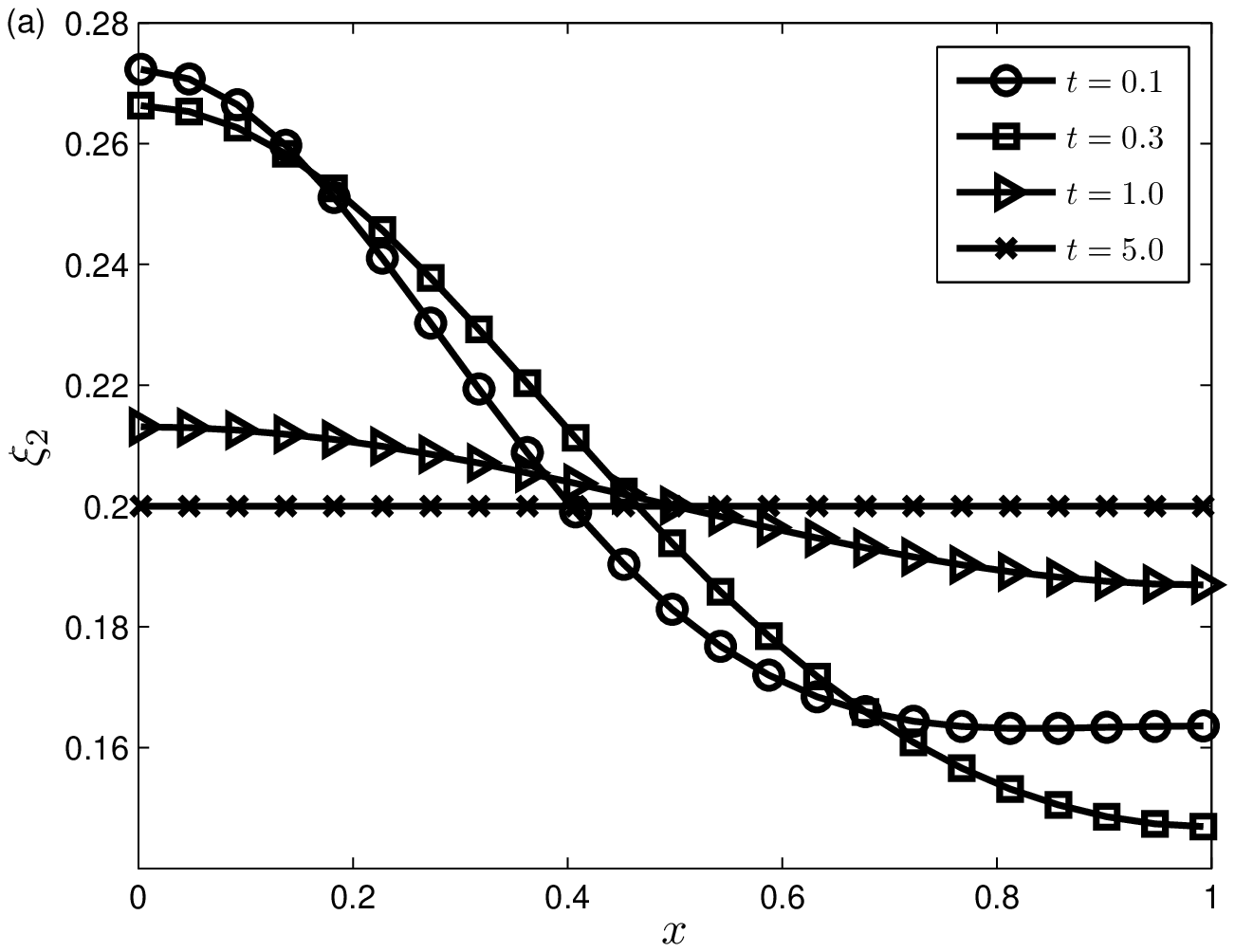}
\includegraphics[width=3.5in]{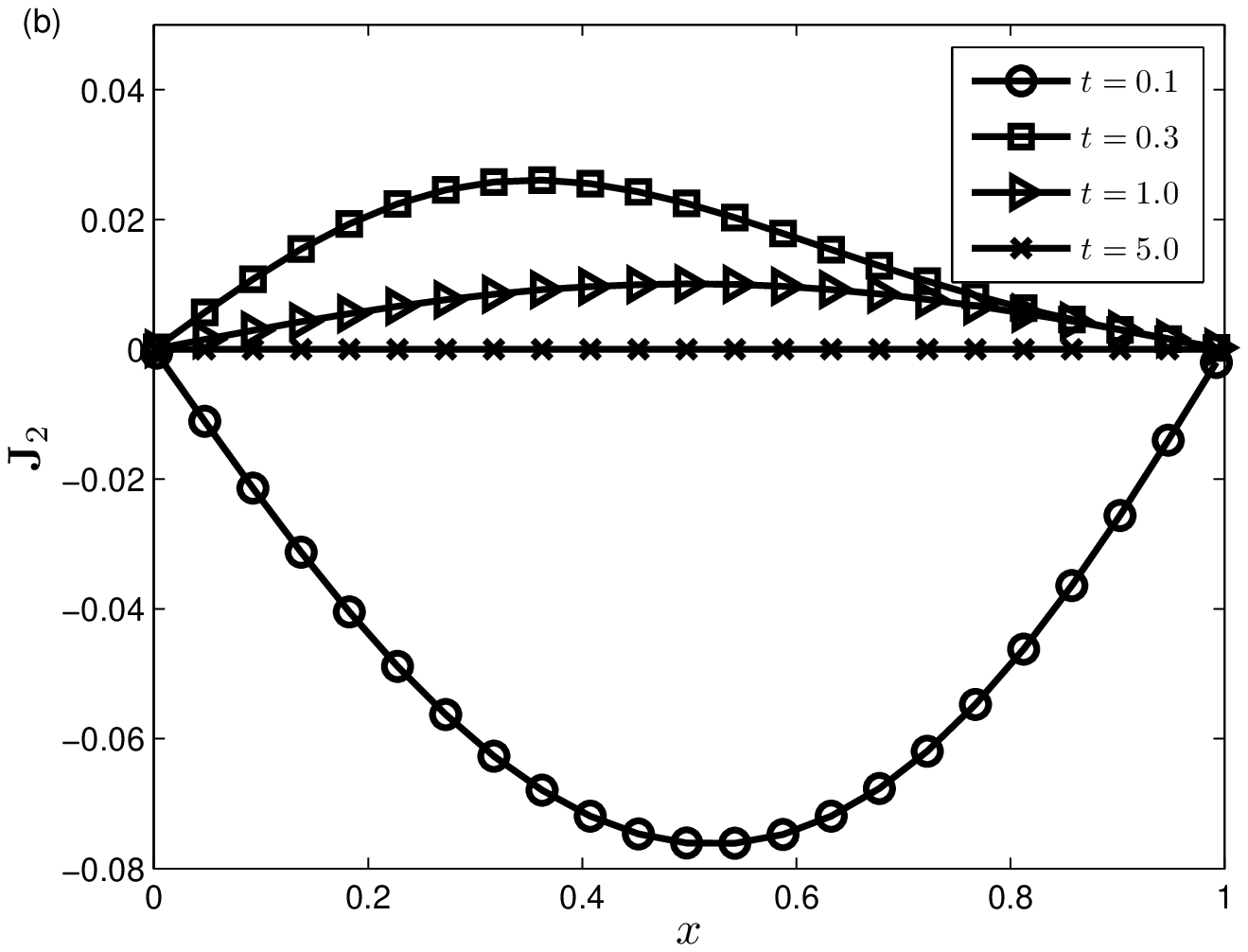}
\includegraphics[width=3.5in]{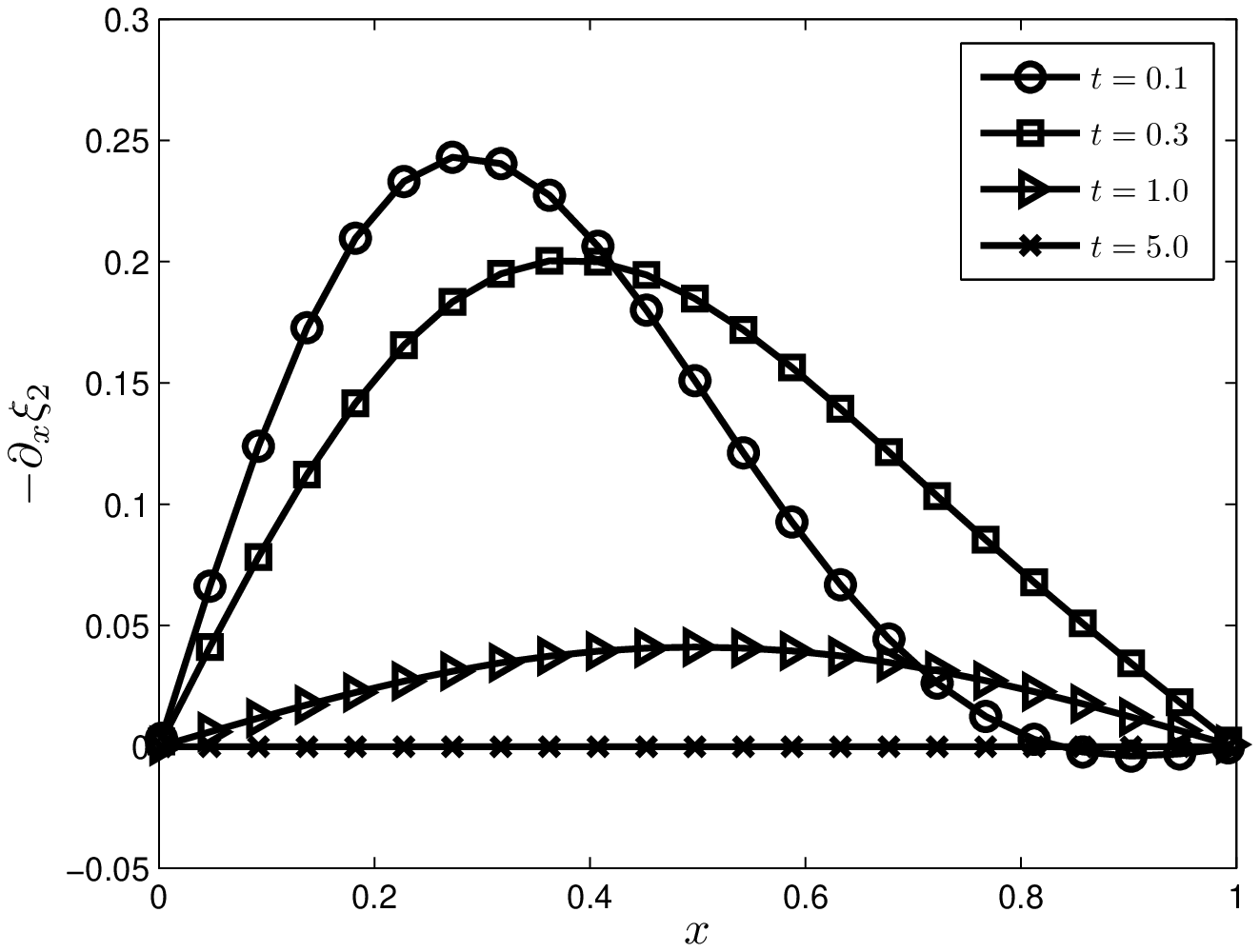}
\centering\caption{\label{fig:9} The distributions of mole fraction $\xi_{2}$, diffusion flux $\mathbf{J}_{2}$ and negative of mole fraction gradient $-\partial_{x}\xi_{2}$ at different time.}
\end{figure}

\subsection{Three-component diffusion in the Loschmidt tube}

Arnold and Toor \cite{Arnold1967} investigated the unsteady diffusion of three components in a Loschmidt tube (see Fig. 10 where the schematic of the problem is presented) with the length ($L$) determined by $(L/\pi)^{2}=1/60$ m$^{2}$, and also found some interesting diffusion phenomena. The system they considered is composed of methane (CH$_{4}$, species 1), argon (Ar, species 2) and hydrogen (H$_{2}$, species 3), and the binary diffusivites among different species are $D_{12}$=21.57 mm$^{2}$/s, $D_{13}$=77.16 mm$^{2}$/s and $D_{23}$=83.35 mm$^{2}$/s \cite{Taylor1993,Arnold1967}. We note that this problem is more complicated than above one since the continuum equations for three species are fully coupled. The initial and boundary conditions of the problem are given by \cite{Taylor1993,Arnold1967}

\begin{subequations}
\begin{equation}
0\leq y \leq L:\ \ \xi_{1}=0.515,\ \xi_{2}=0.485, \ \xi_{3}=0.0,
\end{equation}
\begin{equation}
-L\leq y \leq 0:\ \ \xi_{1}=0.0,\ \xi_{2}=0.509, \ \xi_{3}=0.491,
\end{equation}
\end{subequations}
\begin{equation}
y=\pm L,\ \ \frac{\partial \xi_{i}}{\partial y}=0 \ (i=1,\ 2,\ 3),
\end{equation}
\begin{equation}
x=\pm L,\ \ \xi_{i}|_{x=-L}=\xi_{i}|_{x=L} \ (i=1,\ 2,\ 3).
\end{equation}

\begin{figure}
\includegraphics[width=3.5in]{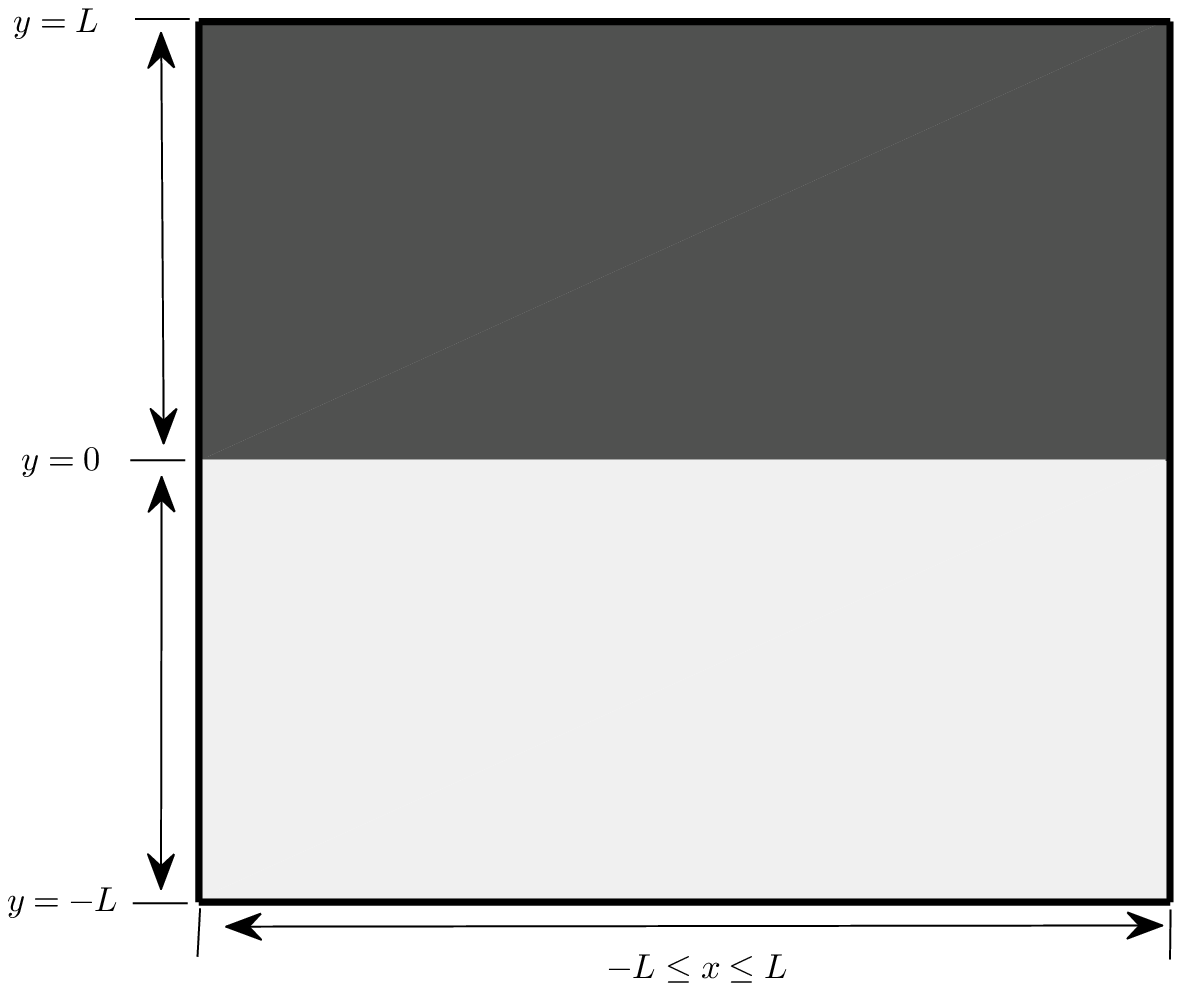}
\centering\caption{\label{fig:10} Schematic of the three-component diffusion in the Loschmidt tube.}
\end{figure}

Before performing simulations, we first introduce the following dimensionless parameters,
\begin{equation}
\bar{x}=\frac{x}{L_{ref}},\ \ \bar{y}=\frac{y}{L_{ref}},\ \ \bar{t}=\frac{t}{t_{ref}},\ \ \bar{D}_{ij}=\frac{D_{ij}t_{ref}}{L_{ref}^{2}},
\end{equation}
where $L_{ref}=L=100\pi\times\sqrt{1/60}$ cm, $t_{ref}=L_{ref}^{2}$ s/cm$^{2}$. Based on above dimensionless parameters, the dimensionless length of Loschmidt tube and the dimensionless diffusivities can be determined as
\begin{equation}
\bar{L}=1.0, \ \ \bar{D}_{12}=0.2157,\ \  \bar{D}_{13}=0.7716, \ \ \bar{D}_{23}=0.8335.
\end{equation}
Similar to above problem, the lattice size $200\times200$ is still applied in our simulations, and to give a comparison between the present results and some available works \cite{Taylor1993,Arnold1967}, here we also measured the average mole fractions $\bar{\xi_{i}}\ (i=1,\ 2)$ in the bottom and top parts of Loschmidt tube,
\begin{equation}
\textrm{Bottom:}\ \bar{\xi_{i}}=\int_{x=-L}^{L}\int_{y=-L}^{0}\xi_{i}dxdy,\ \ \textrm{Top:}\ \bar{\xi_{i}}=\int_{x=-L}^{L}\int_{y=0}^{L}\xi_{i}dxdy,
\end{equation}
and presented the results in Figs. 11 and 12. From these two figures, one can observe that our results are in agreement with the available experimental data and linearized theory \cite{Taylor1993}.
\begin{figure}
\includegraphics[width=3.5in]{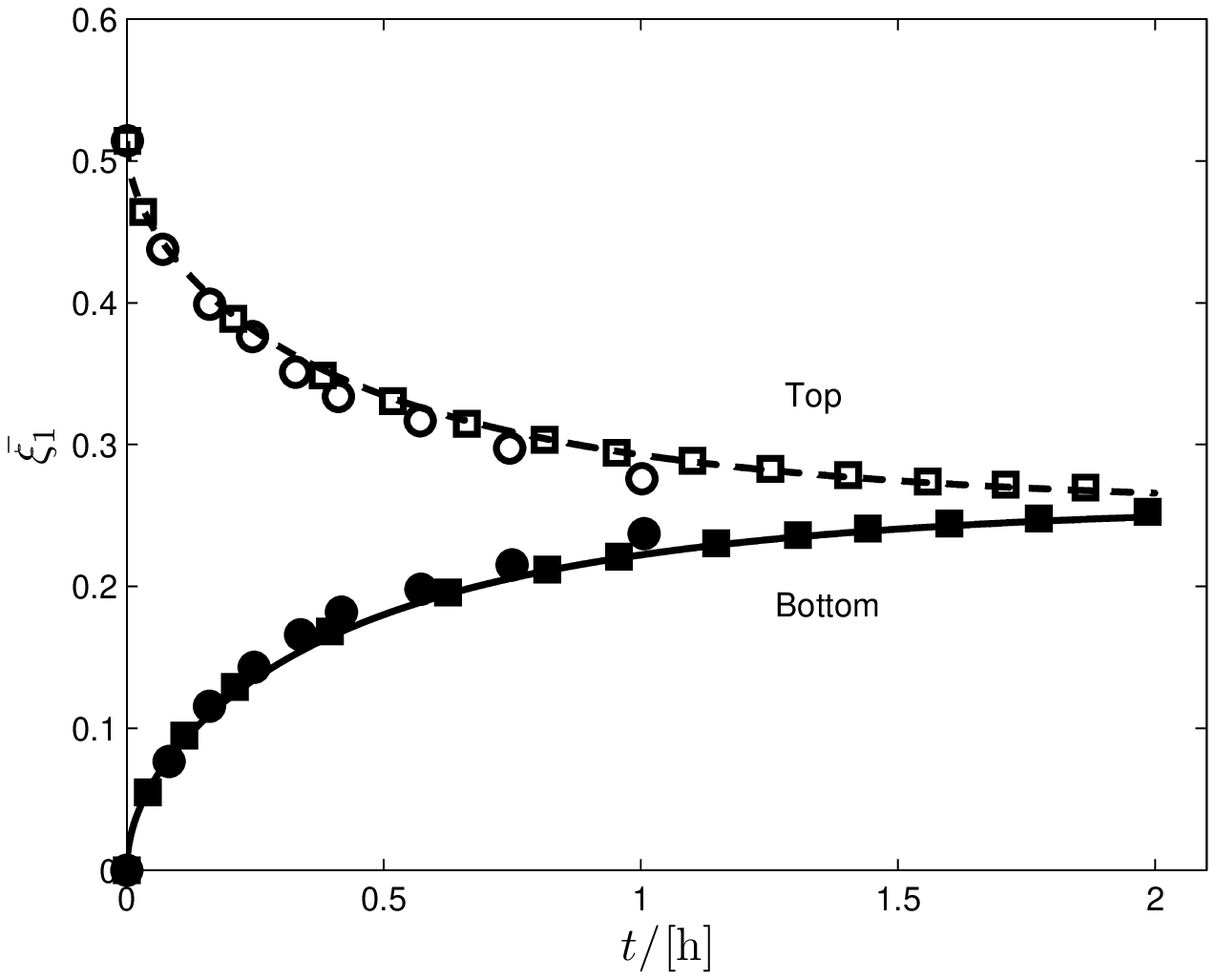}
\centering\caption{\label{fig:11} The average mole fraction $\bar{\xi_{1}}$ at different time (Solid and dashed lines: Present results, $\bigcirc$ and $\vcenter{\hbox{\LARGE$\bullet$}}$: Experimental data \cite{Taylor1993}, $\blacksquare$ and $\square$: Linearized theory \cite{Taylor1993}; h:\ hour).}
\end{figure}
\begin{figure}
\includegraphics[width=3.5in]{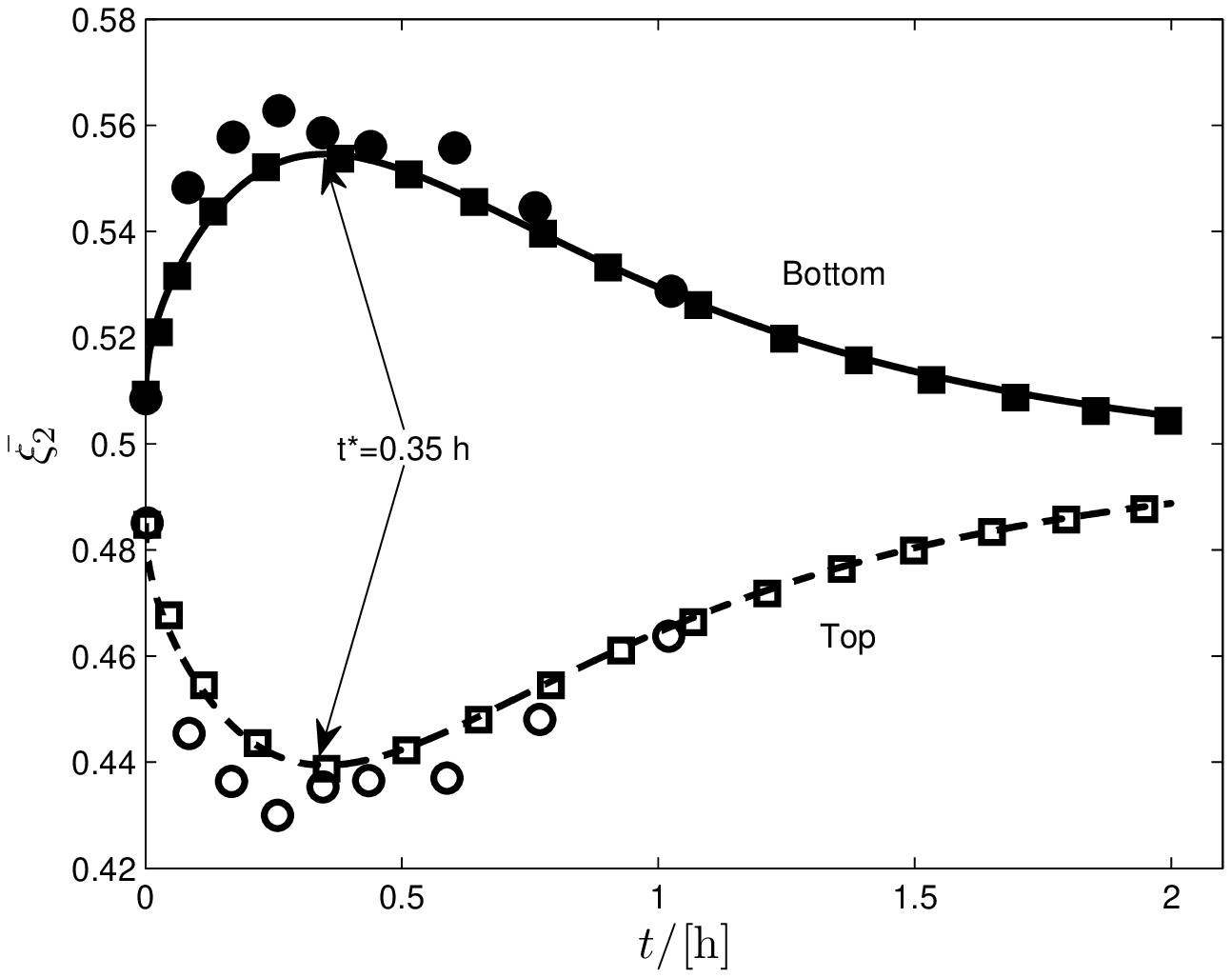}
\centering\caption{\label{fig:12} The average mole fraction $\bar{\xi_{2}}$ at different time (Solid and dashed lines: Present results, $\bigcirc$ and $\vcenter{\hbox{\LARGE$\bullet$}}$: Experimental data \cite{Taylor1993}, $\blacksquare$ and $\square$: Linearized theory \cite{Taylor1993}).}
\end{figure}

In addition, it is also found from Fig.11 that the average mole fraction $\bar{\xi_{1}}$ in top part of the Loschmidt tube decreases with the increase of time, while the average mole fraction $\bar{\xi_{1}}$ in bottom tube shows an opposite trend. Actually, if the time is large enough (e.g., $t=5.0$ h, h denotes the word 'hour'), the average values of mole fraction $\bar{\xi_{1}}$  in the bottom and top parts of the Loschmidt tube would reach to its equilibrium value $\xi_{1}^{*}=0.2575$, which can be seen clearly from the results in Fig. 13(a). We also noted that although there are cross effects for the mole faction $\xi_{1}$, the changes of its average values are similar to theory based on the Fick's law.

Fig. 12 shows the change of the average mole fraction $\bar{\xi_{2}}$ in time. As seen from this figure, the average mole fraction $\bar{\xi_{2}}$ in the bottom part of the Loschmidt tube first increases, and is up to the maximum value at the time $t*=0.35$ h, then it begins to decrease, and would reach to its equilibrium value $\xi_{2}^{*}=0.497$ when the time is large enough (see Fig. 14). However, the average mole fraction $\bar{\xi_{2}}$ in the top part of the Loschmidt tube presents an opposite trend during the time evolution, namely, it first decreases before $t=t*$, then begins to increase when $t>t*$, and approaches to its equilibrium value $\xi_{2}^{*}=0.497$ as time goes on. Compared to the average mole fraction $\bar{\xi_{1}}$, these curious results of the average mole fraction $\bar{\xi_{2}}$ are caused by the cross effects among different species, which can be confirmed by Eq.~(\ref{eq2-13}).

In addition, we also presented the profiles of mole fractions $\xi_{1}$ and $\xi_{2}$ along $y$ direction in Figs. 13 and 14. Similar to the results in Figs. 11 and 12, the mole fraction $\xi_{1}$ in top part of the Loschmidt tube decreases with the increase of time, while the mole fraction $\xi_{1}$ in bottom part of the Loschmidt tube increases in time, and finally both of them reach to the equilibrium value $\xi_{1}^{*}=0.2575$ (see Fig. 13). However, the mole fraction $\xi_{2}$ shows some curious results although it approaches to the equilibrium value $\xi_{2}^{*}=0.497$ with the increase of time. At the beginning, the distribution of mole fraction $\xi_{2}$ in the Loschmidt tube is not far from its equilibrium state ($\xi_{2}^{*}=0.497$), while under the cross effect caused by other species, the larger mole fraction $\xi_{2}$ in the bottom part of the Loschmidt tube further increases, and simultaneously, the smaller mole fraction $\xi_{2}$ in top part of the Loschmidt tube oppositely decreases when the time is less than a critical value (see the results at $t=0.1$ h, 0.3 h in Fig. 14). Then the mole fraction $\xi_{2}$ in bottom part of the Loschmidt tube begins to decrease, and the mole fraction $\xi_{2}$ in top part of the Loschmidt tube increases, and finally they would reach to the equilibrium value $\xi_{2}^{*}=0.497$ (see the results at $t=5.0$ h in Fig. 14).

\begin{figure}
\includegraphics[width=3.5in]{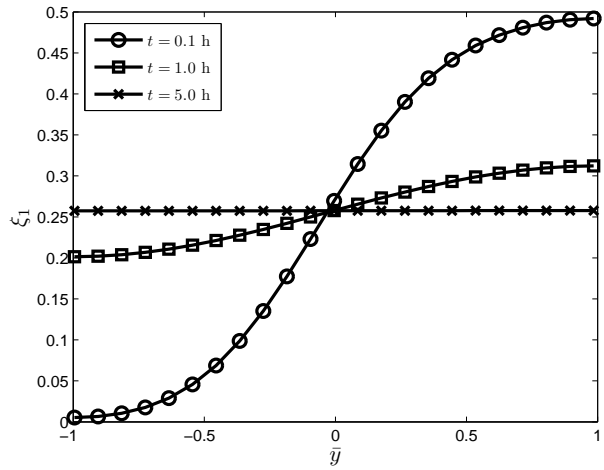}
\centering\caption{\label{fig:13} The profiles of mole fraction $\bar{\xi_{1}}$ along $y$ direction.}
\end{figure}

\begin{figure}
\includegraphics[width=3.5in]{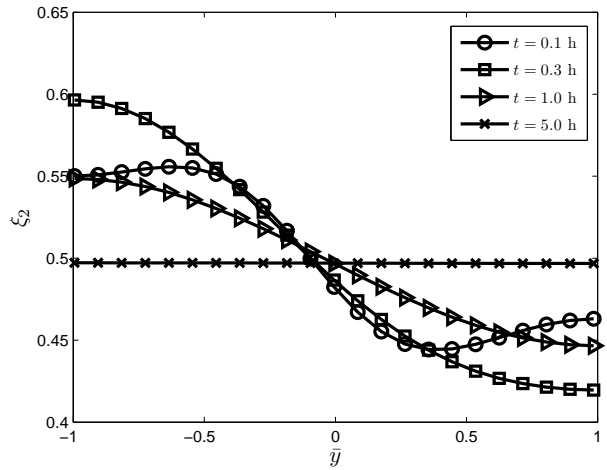}
\centering\caption{\label{fig:14} The profiles of mole fraction $\bar{\xi_{2}}$ along $y$ direction.}
\end{figure}

\section{Conclusions}

In this work, we first developed a MS theory based MRT-LB model for the diffusion in multicomponent mixtures, and also performed a Chapman-Enskog analysis to show that the MS theory based continuum equations can be correctly recovered from present MRT-LB model. Compared to the available LB models based on kinetic theory, the present LB model is much simpler, and does not need to apply any interpolations or finite-difference techniques for the multicomponent diffusion problems with different molecular weights. Then we also tested the developed LB model with some benchmark problems, and found the present results agree well with the analytical solutions, available numerical solutions, the experimental data and/or the approximated linear theory. Besides, we would also like to emphasize that the present LB model can also accurately capture the interesting diffusion phenomena (osmotic diffusion, reverse diffusion and diffusion barrier) inherent in the multicomponennt systems.

Finally, it should be noted that in this work, we only consider the diffusion process in the multicomponennt system. In reality, however, the convection (including diffusion and advection) process, as one of the major types of mass transfer, is more prevalent, and would be investigated in a future work.

\section*{Acknowledgments}

This work was financially supported by the National Natural Science Foundation of China (Grant Nos. 51576079 and 51776068) and the National Key Research and Development Program of China (Grant No. 2017YFE0100100).

\end{document}